\documentclass[aps,prl,reprint,superscriptaddress,nofootinbib,longbibliography,floatfix]{revtex4-2}
\usepackage{amsthm,amsmath,amssymb}
\usepackage{mathrsfs}
\usepackage{graphicx}
\usepackage{array}
\usepackage{float}
\usepackage{CJKutf8}
\newcounter{revpanel}[figure]
\renewcommand{\therevpanel}{\alph{revpanel}}

\newcommand{\revsubfigure}[2]{%
    \begin{minipage}[b]{0.3\textwidth}
        \centering
        #1\par
        \refstepcounter{revpanel}%
        \label{#2}%
        \vspace{1mm}
        (\therevpanel)
    \end{minipage}%
}
\usepackage{dcolumn}
\usepackage{bm}
\usepackage{amsfonts}
\usepackage{xcolor}
\usepackage{appendix}
\usepackage{tabularx}
\usepackage{braket}

\newcommand{\tcell}[1]{\parbox[t]{0.30\textwidth}{\raggedright #1}}
\usepackage[
  colorlinks=true,
  linkcolor=blue,
  citecolor=blue,
  urlcolor=blue
]{hyperref}

\begin{document}
\title{Reliability–Safety Trade-off in AI Distillation: A Renormalization-Group Approach}
\author{Y. M. Du}
\altaffiliation{These authors contributed equally to this work.}
\author{Miao-Miao Yi}
\altaffiliation{These authors contributed equally to this work.}
\author{Tan-Ji Zhou}
\author{C. P. Sun}
\email{suncp@gscaep.ac.cn}
\affiliation{Graduate School of China Academy of Engineering Physics, Beijing 100193, China}

\begin{abstract}
Knowledge distillation transfers more than task competence: it also transmits response propensities, refusal policies, error boundaries, and latent safety biases. We formulate this behavioral inheritance as a coarse-graining model grounded in statistical mechanics, in which the student’s answer and refusal decisions define two macrostates, while the teacher induces an effective field that reshapes the student’s free-energy landscape. The model yields a reliability–safety trade-off relation controlled by a single parameter $K$, which we term the hazard discrimination capability. The predicted trade-off is consistent with refusal-token data [arXiv: 2412.06748]. In knowledge distillation, a teacher with strong hazard discrimination improves the student's attainable reliability and safety, whereas poor discrimination limits the attainable trade-off. Repeated distillation acts as an iterated renormalization-group-like transformation, under which $K$ follows a flow across generations.The flow exhibits a tricritical structure separating regimes of $K$ loss, stable transmission, and threshold-dependent inheritance, and yields testable scaling predictions for multigenerational distillation.
\end{abstract}

\maketitle

\textit{Introduction}.---Some artificial intelligence (AI) systems rely extensively on knowledge distillation during training, compression, and deployment \cite{bucila2006model,ba2014deep,hinton2015distilling,gou2021knowledge,menon2021statistical,stanton2021does}. A compact student model is trained on outputs generated by a larger teacher model; for large language models (LLM), the transferred signals can include class probabilities, token-level distributions, preferences, reasoning patterns, and safety policies \cite{kim2016sequence,sanh2019distilbert,jiao2020tinybert,wang2020minilm,xu2024survey,kodistillm2024distillm,hsieh2023distilling,li2023symbolic,busbridge2025distillation}. Distillation is therefore not merely a compression procedure. It also transfers behavioral decision boundaries \cite{askell2021general,bai2022helpful,ouyang2022training,bai2022constitutional,rafailov2023direct,casper2023open}: later generations may inherit overconfidence, pathological refusal patterns, or a displaced boundary between ordinary and hazardous requests \cite{goodfellow2014explaining,amodei2016concrete,hendrycks2021unsolved,weidinger2021ethical,russell2019human,shumailov2024ai,cloud2026language}. The central question is how a student jointly inherits the teacher's reliability on ordinary inputs and its safety on risky inputs.

From a statistical-physics viewpoint, distillation is closely analogous to coarse-graining \cite{kadanoff1966scaling,wilson1975renormalization} in renormalization-group theory. Both operations reduce internal degrees of freedom while attempting to preserve the desired behavior \cite{tishby1999information,tishby2015deep}. In distillation, a smaller student model is aligned with desired behaviors of a larger teacher model; in renormalization, microscopic variables are eliminated while the long-wavelength behavior is preserved. This analogy motivates a minimal coarse-graining description of AI reliability and safety, in which the macroscopic states are the answer and refusal responses, characterized by their input-dependent free energies \cite{kochjanusz2018mutual,gordon2021relevance}.

We address this question with a two-state (\textit{binary}) model. To
capture the essential physics, we coarse grain the input into ordinary and hazardous classes and the output into answer and refusal macrostates. A Boltzmann-machine
representation is then introduced, with response probabilities
governed by the free energies \cite{shannon1948mathematical,jaynes1957information,mezard2009information,zdeborova2016statistical}
of the answer and refusal states, thereby
revealing an intrinsic reliability–safety trade-off characterized
by the hazard discrimination capacity (HDC). Turning to knowledge
distillation, summing over the teacher's response variables produces an effective external field acting on the student's free-energy landscape. Under recursive distillation, the HDC obeys an renormalization-group like (RG-like) flow governed by the competition between teacher-induced reinforcement and student-capacity loss. A tricritical point organizes the transitions among HDC erasure, persistent transmission, and retention only when the initial teacher's  HDC exceeds a threshold.

\textit{Binary model for the reliability--safety trade-off}.---We
first consider a binary model. Let the coarse-grained input be $x\in\{1,-1\}$, where $x=1$ denotes an ordinary task input and $x=-1$ denotes a hazard input, such as privacy-sensitive or harmful requests. The response is $y\in\{1,-1\}$, with $y=1$ denoting an answer, disclosure, or execution, and $y=-1$ denoting refusal, redaction, clarification, or another conservative response. Reliability and safety are characterized respectively by
\begin{equation}
R=P(y=1|x=1),\qquad S=P(y=-1|x=-1).
\label{eq:RSdef}
\end{equation}
where $R$ is the probability that the model performs the
task on an ordinary input, and $S$ is the probability
that it refuses or adopts a safe response to a hazardous
input.

We introduce a Boltzmann-machine to describe these response probabilities.
Besides the coarse-grained input and output variables, $x$, $y$, let $h$ collect all residual degrees of freedom, including hidden-layer variables in deep levels. For a fixed input $x$, the conditional distribution is taken to be canonical,
\begin{equation}
P(y,h\mid x)=\frac{\exp[-\beta E(y,h;x)]}{\sum_{y',h'}\exp[-\beta E(y',h';x)]},
\label{eq:boltzmann}
\end{equation}
where $E(y,h;x)$ is an energy function describing the model state for input $x$, and $\beta$ is an inverse temperature controlling the stochasticity of the training-induced response distribution.

Since we are concerned only with whether the model ultimately answers or refuses, rather than with the internal degrees of freedom $h$, we integrate out $h$ and thus obtain the reduced conditional
distribution
\begin{equation}
P(y\mid x)
=
\frac{\exp[-\beta F(y,x)]}
{\sum_{y'=\pm1}\exp[-\beta F(y',x)]},
\label{eq:reduced_distribution}
\end{equation}
where
\begin{equation}
F(y,x)
=
-\frac{1}{\beta}
\ln\sum_h \exp[-\beta E(y,h;x)]
\label{eq:freeenergy}
\end{equation}
is the reduced free energy associated with response $y$ for
input $x$. Defining
\begin{equation}
A=\frac{1}{2}\sum_{x,y=\pm1}yF(y,x),
\quad
Q=-\frac{1}{2}\sum_{x,y=\pm1}xyF(y,x),
\label{eq:AQ}
\end{equation}
therefore, the probability of answering input $x$ becomes
\begin{equation}
P(1\mid x)
=
\frac{1}{1+\exp[\beta(A-Qx)]}.
\label{eq:answerprob}
\end{equation}
Here, $A$ characterizes the overall answer-refusal tendency,
increasing $A$ makes the model more inclined to refuse. $Q$ characterizes the alignment between the
input class and the response choice. For $Q>0$, increasing $Q$
raises the answer probability for ordinary inputs and lowers it
for hazardous inputs. For $Q=0$, the model has identical
answer-refusal preferences for the two input classes. whereas $Q<0$ corresponds to the reversed response
pattern.

A reliability–safety trade-off relation for the binary model
\begin{eqnarray}
\mathrm{logit}(R)+\mathrm{logit} (S)=K
\label{eq:tradeoff}
\end{eqnarray}
follows from Eq.~(\ref{eq:RSdef}). Here $K\equiv 2\beta Q$, and $\mathrm{logit} (u)\equiv \ln [u/(1-u)]$. Accordingly, like $Q$, $K$ quantifies the model's ability to distinguish the two input
classes and select the corresponding responses. We refer to
$K$ as the hazard discrimination capability (HDC). Increasing
positive $K$ improves both reliability and safety, thereby
expanding the attainable reliability–safety boundary.

It follows from Eq.~(\ref{eq:RSdef}) that the overall tendency $A$ independently controls the balance between reliability and safety through
\begin{eqnarray}
\mathrm{logit}(S)-\mathrm{logit} (R)=2\beta A.
\label{eq:overall}
\end{eqnarray}
Thus, positive $A$ favors safety at the expense of reliability,
whereas negative $A$ favors reliability at the expense of
safety.

Trade-offs between two competing objectives often share similar mechanistic origins: the same driving force that enhances useful output can also amplify undesirable effects. Like the power-efficiency trade-off in finite-time thermodynamics arises because increasing the thermodynamic driving force can simultaneously raise output power and irreversible dissipation~\cite{ma1,ma2,ma3,ma4}, the reliability–safety trade-off of AI arises because strengthening a model's tendency to refuse $A$ must increases its response probabilities for both ordinary and hazardous inputs.

The overall tendency $A$ can be controlled by adjusting the global answer--refusal threshold, as demonstrated in the experiments of Ref.~\cite{jain2024refusal}. Therefore, these experimental data can provide an empirical check of of the reliability--safety trade-off relation, i.e., Eq.~(\ref{eq:tradeoff}). This prediction is consistent with the refusal-token data of Ref.~\cite{jain2024refusal}: for a fixed model, varying the answer--refusal threshold changes the model's reliability $R$ and safety $S$, while the extracted HDC remains approximately unchanged. As shown in Fig.~2, the data for the ``model without contrast training'' are extracted from Ref.~\cite{jain2024refusal}. The results show that the values of $(R,S)$ corresponding to different thresholds approximately satisfy Eq.~(\ref{eq:tradeoff}), with the extracted HDC values clustering around $\bar K=3.471$. More detailed discussions and additional data plots are provided in Sec.III of Supplemental Material.

\begin{figure}
    \centering
    \includegraphics[width=0.7\linewidth]{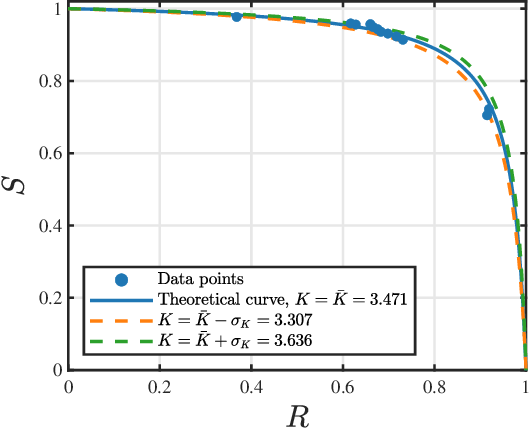}
    \caption{$S$--$R$ plot of the data measured for the ``model without contrast training'' in Ref.~\cite{jain2024refusal} under different answer--refusal thresholds. The data are extracted from Fig.~2 of Ref.~\cite{jain2024refusal} (see Table~S2 in the Supplemental Material) and converted into the corresponding reliability $R$ and safety $S$ according to the definitions in this work. The solid line represents the reliability--safety trade-off curve with the averaged HDC $\bar K=3.471$ obtained from all data points. The two dashed lines represent the trade-off curves with $K=\bar K\pm\sigma_K$, where $\sigma_K=0.164$ is the standard deviation of the HDC extracted from different experimental points.}
    \label{figure_data}
\end{figure}

\textit{Teacher-induced field for student reliability–safety trade-off}---Distillation introduces two binary responses, $y_\mathrm{T}$ for the teacher and $y_\mathrm{S}$ for the student, with $+1$ denoting answer and $-1$ denoting refusal.
To model the knowledge distillation, a teacher-student alignment $V=-\gamma y_\mathrm{S} y_\mathrm{T}$ is introduced, so that the joint conditional distribution is
\begin{equation}
P(y_\mathrm{T},y_\mathrm{S}\mid x)\propto
\exp\!\left[-\beta F_\mathrm{S}(y_\mathrm{S},x)-\beta F_\mathrm{T}(y_\mathrm{T},x)+\gamma y_\mathrm{S}y_\mathrm{T}\right],
\label{eq:coupled}
\end{equation}
where $F_\mathrm{T}$ and $F_\mathrm{S}$ are the uncoupled teacher and pre-distillation student free energies, respectively.

The post-distillation student can be described by the reduced conditional distribution
\begin{equation}\label{eq9}
P_\mathrm{S}^{\mathrm{eff}} (y_\mathrm{S}|x)=\frac{\exp[-\beta F_\mathrm{S}^{\mathrm{eff}}(y_\mathrm{S},x)]}{\sum_{y_s}\exp[-\beta F_\mathrm{S}^{\mathrm{eff}}(y_\mathrm{S},x)]},
\end{equation}
where the effective free energy after distillation is
\begin{equation}
    F_\mathrm{S}^{\mathrm{eff}}(y_\mathrm{S},x)=F_\mathrm{S}(y_\mathrm{S},x)-\frac{y_\mathrm{S}}{2\beta}B_{\mathrm{T}}(x).
\end{equation}
with
\begin{equation}
B_\mathrm{T} (x)=2\mathrm{arctanh}[m_\mathrm{T} (x)\tanh\gamma],
\end{equation}
and $m_\mathrm{T} (x)=P_\mathrm{T} (1|x)-P_\mathrm{T} (-1|x)$. Here, the ``magnetization'' $m_\mathrm{T}(x)$ characterizes the teacher's response bias for input $x$. The role of the teacher is therefore compressed into an input-dependent external field $B_\mathrm{T}(x)$ that reshapes the student's answer-refusal free-energy landscape.

Accordingly, the student's HDC is also renormalized by the teacher-induced field $B_\mathrm{T}(x)$. Let $K_0$ be the initial student's HDC before distillation.
After distillation, the effective HDC
becomes
\begin{equation}
K_{\mathrm{eff}}=K_0+B_\mathrm{T}(1)-B_\mathrm{T}(-1),
\label{eq:Keffdef}
\end{equation}
for the post-distillation reliability–safety trade-off is
$
\mathrm{logit}R_\mathrm{S}^{\mathrm{eff}}+\mathrm{logit} S_\mathrm{S}^{\mathrm{eff}}=K_{\mathrm{eff}}$.

\begin{figure}[htbp]
	\centering	

	\includegraphics[scale=0.3]{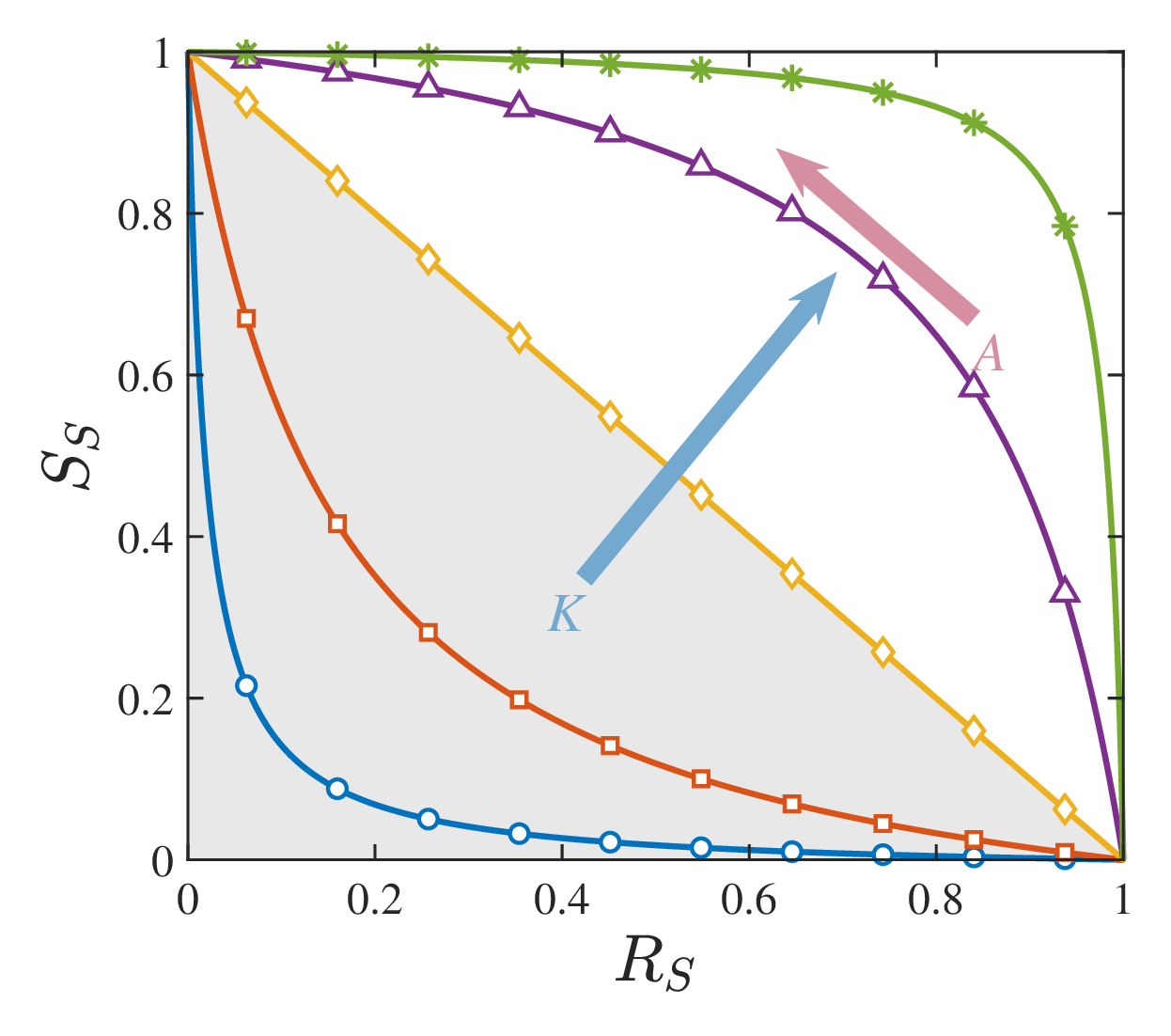}
	\caption{Reliability--safety trade-off curves for different effective HDC values $K_{\mathrm{eff}}$: $K_{\mathrm{eff}}=-4$ (blue circles), $-2$ (red squares), $0$ (yellow diamonds), $2$ (purple upward triangles), and $4$ (green star). The gray shaded region corresponds to $K_{\mathrm{eff}}<0$, or equivalently $R+S<1$, and is bounded by the $K_{\mathrm{eff}}=0$ line. The light-blue and light-pink arrows indicate the directions of increasing \(K_{\mathrm{eff}}\) and \(A_{\mathrm{eff}}\), respectively.}\label{figure0}
\end{figure}

If the teacher tends to answer ordinary inputs and refuse hazardous ones, then $B_\mathrm{T}(1)>B_\mathrm{T}(-1)$. In this case, distillation increases the student's $K_\mathrm{eff}$, thereby improving the attainable reliability–safety trade-off, as shown in Fig.~ \ref{figure0}(a). Conversely, if the teacher also exhibits a strong tendency to answer hazardous inputs, the field contrast between the two input classes may be reduced or even reversed, thereby lowering $K_\mathrm{eff}$ and worsening its trade-off. Therefore, the crucial element determining the student’s post-distillation reliability–safety is the teacher-induced field contrast across input classes, rather than the teacher’s task accuracy alone.

Let $A_{0}$ be the overall tendency to refuse of the pre-distillation student model.
The effective overall tendency after distillation
becomes
\begin{equation}
A_{\mathrm{eff}}=A_0-\frac{1}{2\beta}[B_\mathrm{T}(1)+B_\mathrm{T}(-1)].
\label{eq:Aeffdef}
\end{equation}
which shows that the teacher's overall response tendency is also transferred to the student.

\textit{Multigenerational distillation}.---We next consider recursive distillation, in which each distilled student becomes the teacher of the next generation.  We focus on the weak-alignment limit $\gamma\ll1$, for which the effective field induced by generation $l$ on generation $l+1$ is $B_l(x)\simeq 2\gamma m_l(x)$.

Let $K_l$ be the HDC of the $l$-th generation and $K^0_{l+1}$ be the pre-distillation HDC of generation $l+1$. We define
\begin{equation}
a_{l}\equiv1-{K_{l+1}^{0}}/{K_{l}},
\end{equation}
which quantifies the relative HDC loss of the pre-distillation student with respect to its teacher.. In the Supplemental Material, through a concrete microscopic model example for hidden layers, we illustrate that the relative HDC loss HDC $a_l$ originates from the reduction of hidden-layer size (i.e., the number of hidden degrees of freedom) in the student model relative to the teacher model. Let $s_l$ denote the relative reduction in model size from the $(l-1)$-th generation to the $l$-th generation. We further derive the relation between $a_l$ and $s_l$. For simplicity, we consider the case where the relative reduction in model size is identical across generations, i.e., $s_l=s$.
Therefore, $a_l$ becomes independent of $l$, and we denote it as
$a_l=a$.

Similarly, let $A_l^0$ and $A_l$ denote the overall tendency to refuse of the $l$-th generation pre- and post-distillation student, and we consider the same pre-distillation tendency for all generations, $A_l^0\equiv A_0$.
For $\gamma\ll 1$, the effective fields generated by generation $l$ for the two input classes satisfy $B_l (1)-B_l (-1)=4\gamma(R_l +S_l-1)$ and $B_l (1)+B_l (-1)=4\gamma(R_l -S_l)$. Consequently, in the continuum limit, we treat the generation index $l$ as a continuous variable, and the distillation dynamics are governed by the RG-like equation
\begin{eqnarray}
\frac{\mathrm{d}K}{\mathrm{d}l}
&=&-aK
+4\gamma\frac{\sinh(K/2)}
{\cosh(K/2)+\cosh(\beta A)},\nonumber\\
\frac{\mathrm{d}A}{\mathrm{d}l}
&=&A_{0}-A
+\frac{2\gamma}{\beta}
\frac{\sinh(\beta A)}
{\cosh(K/2)+\cosh(\beta A)}.
\label{eq:iteration}
\end{eqnarray}
Although we derived these equations under specific conditions for $A_0$ and $K_0$, as discussed below, the phase structure predicted by this RG-like flow persists within a broader class of inheritance dynamics.

\textit{RG-like flow for AI}.---
The analytical treatment of the RG-like equations in Eq.~(\ref{eq:iteration}) is provided in the supplementary material; here, we present only the main conclusions and the numerical results.
The fixed points of the RG-like equations depend only on two independent parameters $|\beta A_{0}|$ and $a/\gamma$. The former characterizes the student's overall response tendency before distillation. When $A_0=0$, the pre-distillation student has equal reliability and safety. The latter is the ratio of the relative HDC loss in the student model to the alignment strength and characterizes the difficulty of transferring capability from the teacher to the student.
\begin{figure*}[t]
    \revsubfigure{%
        \includegraphics[scale=0.3]{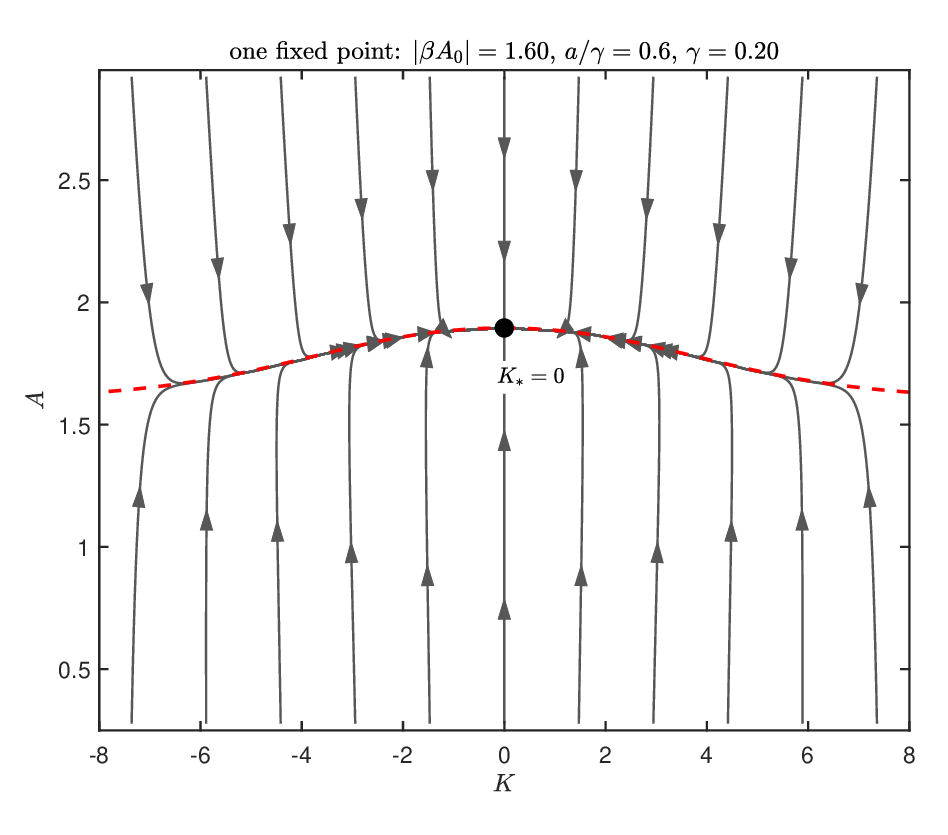}%
    }{fig:figure1a}
    \hspace{0\textwidth}
    \revsubfigure{%
        \includegraphics[scale=0.3]{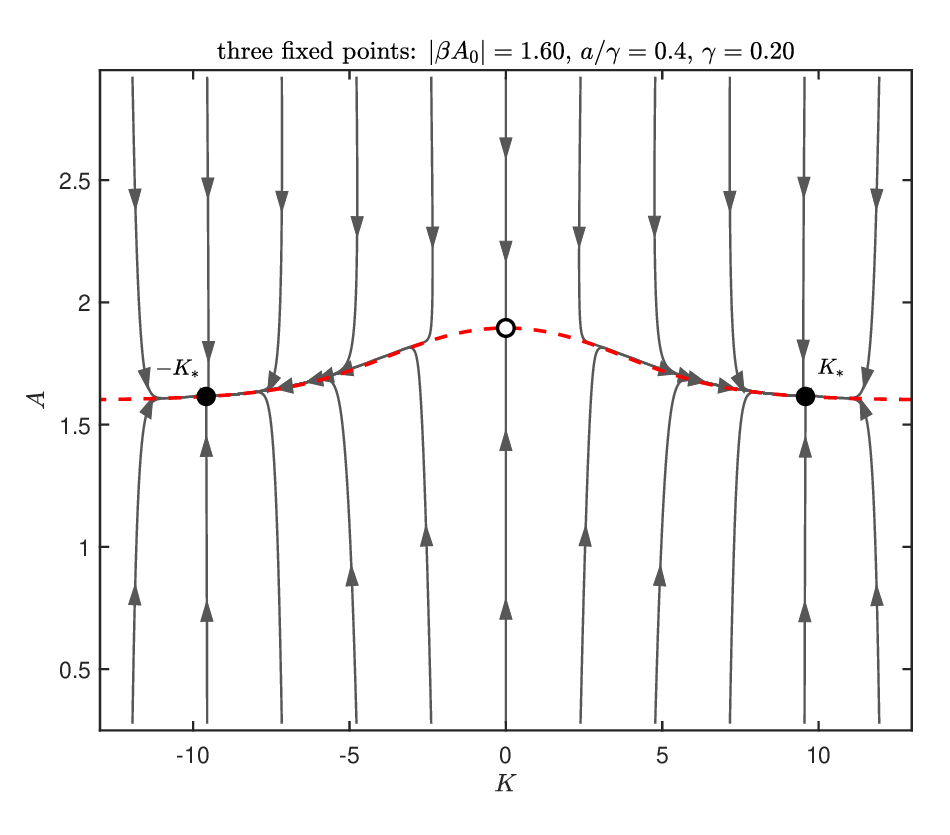}%
    }{fig:figure1c}
   \hspace{0\textwidth}
    \revsubfigure{%
        \includegraphics[scale=0.3]{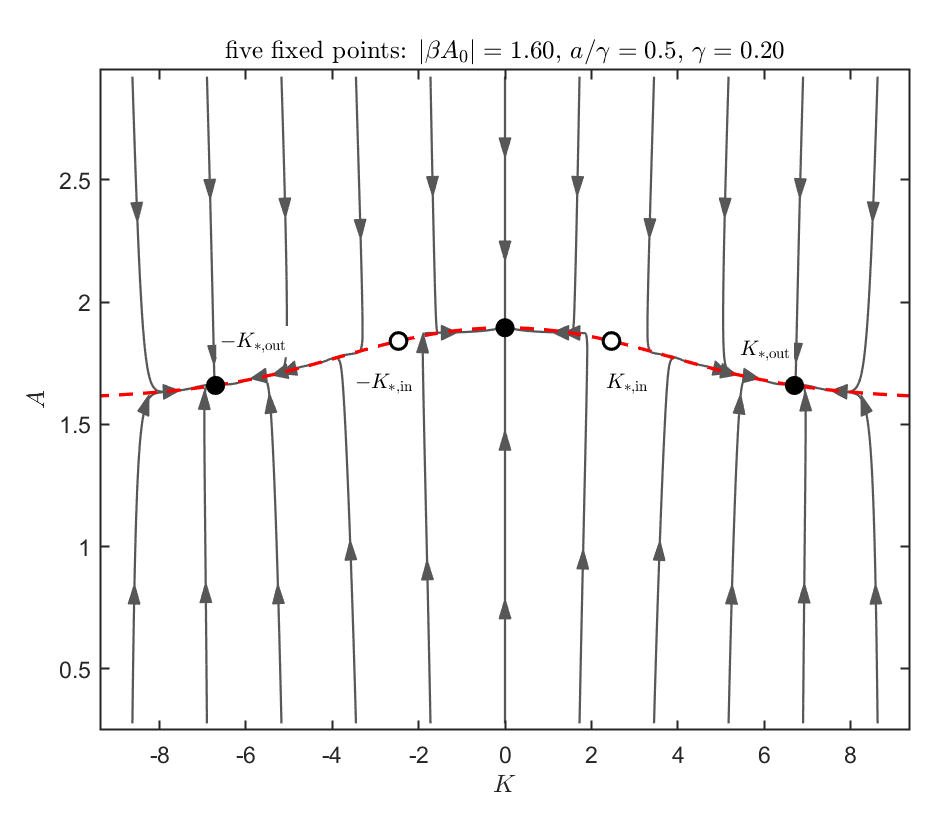}%
    }{fig:figure1b}
     \caption{RG flows for the three regimes: (a) regime I (b) regime II (c) regime III. The black and white dots denote the stable and unstable fixed points, respectively. The arrows represent the direction of the flows. The red dash curve shows the stable condition  in $A$ direction $\mathrm{d}A/\mathrm{d} l=0$. }
    \label{figure2}
\end{figure*}

We next analyze the fixed-point configurations of Eqs.~(19) in different regions of the $(|\beta A_0|,a/\gamma)$ parameter space. The corresponding RG-like flows are shown in Fig. ~\ref{figure2}. Owing to the $Z_2$ symmetry of the equation $K\mapsto-K$, nonzero fixed points occur in symmetry-related pairs.
Away from the bifurcation boundaries, the RG flow exhibits three generic fixed-point configurations, containing one, three, and five distinct real fixed points, respectively. This implies an underlying $Z_2$-symmetric Ginzburg-Landau theory. 

In the weak-alignment regime, $A$ rapidly approaches the stationary state satisfying $\mathrm{d}A/\mathrm{d}l=0$, whereas $K$ evolves more slowly, as shown in Fig.~\ref{figure2}. On the scale over which $K$ changes appreciably, the stationary solution ($\mathrm{d}A/\mathrm{d}l=0$) for $A$ can therefore be substituted into the $K$ flow equation in Eqs.~(18), reducing the two-dimensional $(K,A)$ flow in Eqs.~(18) to the following one-dimensional Ginzburg-Landau expansion,
\begin{equation}\label{eq:landau_expansion}
\begin{split}
\frac{\mathrm{d}K}{\mathrm{d}l}\simeq&\,\mu K+\frac{\gamma(C-2)}{12(1+C)^{2}}K^3\\
&\,+\frac{\gamma(C^2-13 C+16)}{960(1+C)^3}K^5,
\end{split}
\end{equation}
with $C=\cosh |\beta A_0|$ and $\mu=2\gamma/(1+C)-a$.

\begin{figure}[htbp]
	\centering	
	\includegraphics[scale=0.38]{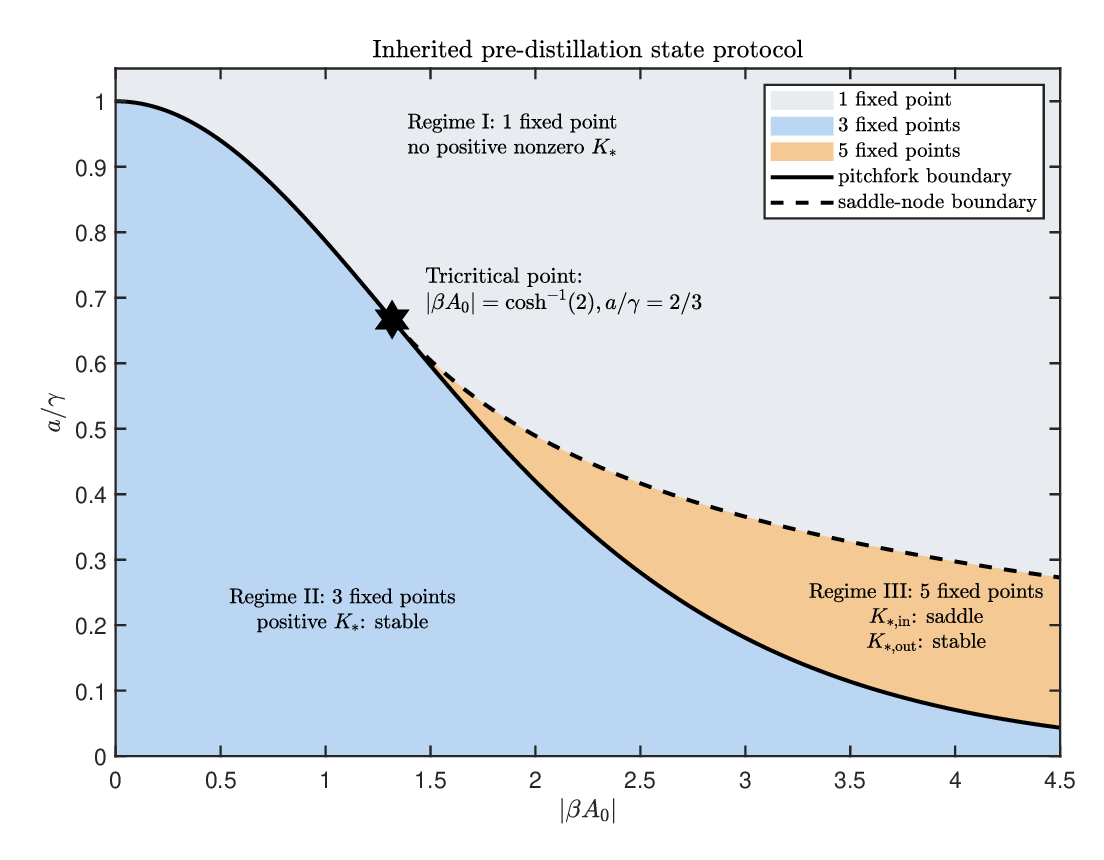}
	\caption{RG fixed-point bifurcation diagram for multigenerational distillation. The gray, blue, and orange regions denote Regimes I, II, and III, containing one, three, and five distinct fixed points, respectively. In Regime III, the central fixed point and the outer pair are stable, whereas the inner pair consists of saddles. The solid curve is the pitchfork bifurcation line $a/\gamma=\operatorname{sech}^{2}(\beta A_{0}/2)$.
The dashed curve is the saddle-node bifurcation line obtained from the maximum of the nonzero fixed-point branch. The black star marks the codimension-two tricritical bifurcation point $(|\beta A_{0}|,a/\gamma)=(\operatorname{arcosh}2,2/3)$.
}
	\label{figure3}
\end{figure}

Accordingly, the number and stability of the real fixed points of this Ginzburg-Landau normal form naturally partition the parameter space into three regimes.
\textit{Regime I.} For $C\leq 2$ and $\mu<0$, there is only one stable fixed point at $K=0$, indicating that the student model is too small to acquire capabilities from the teacher model (see Fig.~\ref{figure2} (a)).
\textit{Regime II.} For $\mu>0$ there are three fixed points that are denoted by $0,\pm K_{*}$, with $K=0$ being unstable and $K=\pm K_*$ being stable, indicating that the student model is sufficiently large to acquire and retain capabilities from the teacher model (see Fig.~\ref{figure2} (b)).
\textit{Regime III.} For $C>2$, the positive cubic term leads to a subcritical bifurcation, while the higher-order nonlinear terms determine the subsequent stabilization and produce a bistable interval on the $\mu<0$ side.
Within this interval, there are five fixed points, $0,\pm K_{*,\mathrm{in}},\pm K_{*,\mathrm{out}}$, with $K=0$ and $K=\pm K_{*,\mathrm{out}}$ being stable and $K=\pm K_{*,\mathrm{in}}$ ($K_{*,\mathrm{out}}>K_{*,\mathrm{in}}$) being unstable, indicating that whether the student model can retain the acquired capabilities depends on whether the HDC of the original teacher $K_{\mathrm{T}}$ exceeds the threshold determined by $K_{*,\mathrm{in}}$ (see Fig.~\ref{figure2} (c)). The two bifurcation sectors meet at
$
C=2,
$
which identifies the tricritical point of the Ginzburg-Landau theory. The RG fixed-point bifurcation diagram of these three regimes is presented in Fig.~\ref{figure3}. A more detailed discussion of the fixed-point structures in different parameter regions of of the phase diagram is provided in the Sec.~II of Supplemental Material.

Notably, the Ginzburg-Landau expansion in Eq.~(\ref{eq:landau_expansion}) requires only that $A$ can be eliminated from the flow. Therefore, although the above discussion focuses on the case of constant $A_0$, it can be extended to more general cases, in which the basic structure of the one-, three-, and five-fixed-point regions and their bifurcation boundaries remains unchanged (see the supplementary material for the precise conditions)

Next, we study the bifurcation scaling of the fixed points. We focus on the scenario in which $|\beta A_0|$ and $a$ are fixed while the alignment strength $\gamma$ is increased. This decreases $a/\gamma$ and moves the system vertically downward in Fig.~\ref{figure3}.

The Regime-I--Regime-II and the Regime-III--Regime-II boundaries are the
pitchfork bifurcations, where the stable nonzero pair
exhibits the square-root onset
\begin{equation}
K_{*}
\propto
|\gamma-\gamma_{\mathrm{PF}}|^{1/2},
\end{equation} where $\gamma_{\mathrm{PF}}=a\cosh^{2}(\beta A_0/2)$ is the bifurcation value.

At the tricritical bifurcation point, the cubic term in Eq.~(\ref{eq:landau_expansion})
vanishes and the direct Regime-I--Regime-II transition
instead exhibits the fourth-root onset
\begin{equation}
K_{*}
\propto
|\gamma-\gamma_{\mathrm{tri}}|^{1/4},
\end{equation}
where the tricritical point $\gamma_{\mathrm{tri}}=3a/2$.
Hence, when the control path is tuned through the tricritical
bifurcation point, the generic square-root scaling of the pitchfork bifurcation is replaced by fourth-root scaling.

The Regime-I--Regime-III boundary is a saddle-node
bifurcation, at which the separation between the inner
saddle and outer stable node obeys
$
K_{*,\mathrm{out}}-K_{*,\mathrm{in}}
\propto
|\gamma-\gamma_{\mathrm{SN}}|^{1/2},
$
where $\gamma_{\mathrm{SN}}$ denotes the corresponding saddle-node bifurcation value.

The theory leads to several direct predictions. First, multigenerational distillation does not improve or degrade the HDC without bound; instead, it saturates at a fixed point. Second, a capacity threshold should be observable. If each student is too small, repeated distillation drives the system to $K_*=0$ fixed point, at which the HDC is lost. Furthermore, increasing the alignment strength $\gamma$ should reduce this minimum-capacity threshold $\gamma_{c}$ whose value depends on the boundary shown in Fig.~\ref{figure3}. In particular, in the low strength condition $\gamma\ll1$,  $\gamma_{c}$ is proportional to the
relative reduction in model size from the teacher to the student $s$,  $\gamma_{c}\propto s$ (see supplementary material).

\textit{Discussion}.---We propose a minimal statistical-physics model of the reliability--safety trade-off and knowledge distillation in AI, centered on the hazard discrimination capability $K$ and directly supported by the refusal-token experiment of Ref.~\cite{jain2024refusal}. The model gives two physical interpretations of distillation: first, free-energy renormalization, whereby the teacher shifts the free-energy describing the student's answer and refusal states; second, in multigenerational distillation, transmission of the macroscopic discrimination capability $K$ across generations, generating an RG-like flow.

Our main result is a phase diagram generated by the competition between teacher-induced reinforcement and student-capacity loss. Its one-, three-, and five-fixed-point regions are separated by pitchfork and saddle-node boundaries meeting at a tricritical point. They correspond, respectively, to hazard discrimination capability loss (Regime~I), stable transmission (Regime~II), and inheritance only above an initial-teacher HDC threshold (Regime~III). This structure also explains the deterioration observed for much smaller students~\cite{busbridge2025distillation,Mirzadeh,Zhang1}: compression removes discrimination information faster than the teacher signal can restore it, preventing the hazard discrimination capability retention.

It is worth noting that the above conclusions, particularly the phase diagram structure, are not limited to the specific case considered in this work, namely, $A_0=\mathrm{const}$ and $K_0=(1-a)K$. The linear inheritance rule may be relaxed to any smooth odd
function $K_0=g(K)$, provided that the reduced $K$ flow retains
a nondegenerate quintic tricritical normal form and develops no
additional extrema at finite $K$. For $A_0=f(A)$ with a fast
attracting fixed point; these generalizations only renormalize the
Ginzburg-Landau coefficients and phase boundaries. Weak $Z_2$ breaking
preserves the hyperbolic fixed-point structure away from the
boundaries but unfolds the symmetric bifurcations, leaving the
tricritical diagram as an organizing skeleton for a broader class
of distillation dynamics.

\textit{Acknowledgment}.---This work was supported by the Science Challenge Project (Grant No. TZ2025017), the National Natural Science Foundation of China (NSFC) (Grant No. 12688201).

%

\clearpage
\onecolumngrid
\setcounter{section}{0}
\setcounter{subsection}{0}
\setcounter{equation}{0}
\setcounter{figure}{0}
\setcounter{table}{0}
\setcounter{secnumdepth}{3}
\renewcommand{\thesection}{\Roman{section}}
\renewcommand{\thesubsection}{\Alph{subsection}}
\renewcommand{\theequation}{S\arabic{equation}}
\renewcommand{\thefigure}{S\arabic{figure}}
\renewcommand{\thetable}{S\arabic{table}}
\makeatletter
\renewcommand{\fnum@figure}{\figurename~\thefigure}
\renewcommand{\fnum@table}{\tablename~\thetable}
\makeatother
\renewcommand{\theHsection}{SM.\Roman{section}}
\renewcommand{\theHsubsection}{SM.\Roman{section}.\Alph{subsection}}
\renewcommand{\theHequation}{SM.\arabic{equation}}
\renewcommand{\theHfigure}{SM.\arabic{figure}}
\renewcommand{\theHtable}{SM.\arabic{table}}

\begin{center}
{\large\bfseries Supplemental Material for ``Reliability–Safety Trade-off in AI Distillation: A Renormalization-Group Approach''\par}
\vspace{1.2em}
Y. M. Du, Miao-Miao Yi, Tan-Ji Zhou, and C. P. Sun\\[0.5em]
Graduate School of China Academy of Engineering Physics, Beijing 100193, China
\end{center}
\vspace{1em}

\section{Relation between the HDC loss parameter $a$ and the model-size reduction $s$: an example based on a microscopic model}
In the main text, we introduced the parameter $a$ to quantify the relative loss of the student's hazard discrimination capability (HDC) before distillation with respect to that of the teacher model. In this section, we use an explicit microscopic model to illustrate intuitively how the parameter $a$ arises from the reduction of hidden degrees of freedom, and to demonstrate the relation between $a$ and the change in hidden-layer size of the student model relative to the teacher model. This example shows that the HDC loss of the student ``small model'' relative to the teacher ``large model'' originates from the reduction of its hidden-layer size relative to that of the teacher model.

\subsection{Derivation of $a$}
We first consider a teacher model with input $x$, response $y$, and internal hidden degrees of freedom $h$. Let $x=\pm1$ denote the input, where $x=+1$ represents an ordinary input and $x=-1$ represents a hazardous input. Let $y=\pm1$ denote the response, where $y=+1$ represents a normal response and $y=-1$ represents refusal. The hidden layer of the teacher model is $h=(h_1,\cdots,h_{N_\mathrm{T}})^T$, where $N_\mathrm{T}$ is the number of hidden degrees of freedom of the teacher model, and each hidden degree of freedom $h_j$ is taken to be a continuous variable. We consider the following energy functional for the teacher model
\begin{equation}
E_\mathrm{T}(y,  h;x)
=
\frac{1}{2}h^{T}Jh
-x u^{T}h
-\frac{y}{2}c^{T}h
+\frac{Ay}{2},  
\qquad
J\succ0.
\end{equation}
Here, $J$ is a positive-definite $N_\mathrm{T}\times N_\mathrm{T}$ matrix characterizing interactions among the hidden degrees of freedom, while $u$ and $c$ describe the couplings between the input and hidden degrees of freedom and between the response and hidden degrees of freedom, respectively.

We next consider the HDC of the teacher model. Define
\begin{equation}
\eta(y,  x)=x u+\frac{y}{2}c.
\end{equation}
The partition function of the teacher model is then
\begin{align}
Z_\mathrm{T}(y;x)
=&\,  \int\mathrm{d}h\exp[-\beta E_\mathrm{T}(y,  h;x)]\notag\\
=&\,  \left(\frac{2\pi}{\beta}\right)^{N_\mathrm{T}/2}
(\det J)^{-1/2}
\exp\left[
-\frac{\beta Ay}{2}
+\frac{\beta}{2}\eta^{T}J^{-1}\eta
\right].
\end{align}
This gives the effective free energy
\begin{equation}
    F_\mathrm{T}(y,  x)
=
-\beta^{-1}\ln Z_\mathrm{T}(y;x)
=\frac{y}{2}(A-xu^TJ^{-1}c)+\textit{const},  
\end{equation}
where the $\textit{const}$ is independent of $x$ and $y$. According to Eq.~(5) of the main text, we have
\begin{equation}
    Q_{\mathrm{T}}=-\frac{1}{2}\sum_{x,  y}xy F_{\mathrm{T}}(y,  x)=u^TJ^{-1}c,  
\end{equation}
and the HDC of the teacher model is
\begin{equation}
    K_{\mathrm{T}}=2\beta Q_{\mathrm{T}}=2\beta u^TJ^{-1}c.
\end{equation}

For a more direct comparison with the teacher model, we further consider a smaller student model. We partition the teacher hidden layer as $h=(z,w)^T$, where $z=(z_1,\cdots,z_{N_\mathrm{S}})^T$ and $w=(w_1,\cdots,w_{N_\mathrm{T}-N_\mathrm{S}})^T$. Correspondingly, the couplings $u$ and $v$ between the input layer, response layer, and hidden layer are decomposed as $u=(u_z,u_w)$ and $v=(v_z,v_w)$, and the interaction matrix among the hidden degrees of freedom is written as
\begin{equation}
J=\begin{pmatrix}J_{zz}&J_{zw}\\J_{wz}&J_{ww}\end{pmatrix}.
\end{equation}
The energy function of the student is taken to be
\begin{equation}
E_{\mathrm{S}}(y,  z;x)
=
\frac{1}{2}z^{T}J_{zz}z
-xu_{z}^{T}z
-\frac{y}{2}c_{z}^{T}z
+\frac{y}{2}(A-bx).
\end{equation}
Here, $z$ denotes the hidden variables of the student model, whose hidden-layer size is $N_\mathrm{S}$ with $N_\mathrm{S}<N_\mathrm{T}$. The matrix $J_{zz}$ is an $N_\mathrm{S}\times N_\mathrm{S}$ matrix, and $u_z$ and $c_z$ are the components of the teacher coupling vectors in the retained subspace.

The HDC of the student model is therefore
\begin{equation}
    K_{\mathrm{S}}^0=2\beta u_z^TJ_z^{-1}c_z.
\end{equation}
According to the definition of the HDC loss parameter $a$ in the main text, one obtains
\begin{equation}
    a=1-\frac{
u_{z}^{T}
J_{zz}^{-1}c_{z}
}{
u^{T}
J^{-1}c
}.
\end{equation}
When the student model contains all hidden degrees of freedom of the teacher model, namely $N_\mathrm{S}=N_\mathrm{T}$, we have $z=h$, $u_z=u$, $J_{zz}=J$, and $v_z=v$, so that $a=0$, corresponding to no HDC loss. This result shows that $a$ directly quantifies the reduction of the reliability-safety trade-off boundary caused by the decrease in the hidden-layer size of the student model.

\subsection{Relation between $a(l)$ and the student-teacher model-size scaling ratio $s_l$ under generational distillation}
We next consider a generational distillation process, in which the model at generation $l$ becomes the teacher for generation $l+1$. Let the number of hidden-layer degrees of freedom of the generation-$l$ model be $N_l$. We define $s_l=1-N_{l+1}/N_l$ as the scale-reduction ratio of the hidden-layer size of generation $l$ relative to the previous generation.

The following explicit example illustrates the relation between $a(l)$ and the scale-reduction ratio $s$. Building on the student model of the previous subsection, we further consider a more specific hidden-layer structure.

First, let the hidden-layer interaction matrix of the generation-$l$ model be $J$, satisfying $Je_q=\lambda_q e_q$, where $\lambda_q$ and $e_q$ denote the $q$th eigenvalue and the corresponding eigenmode, respectively. We order them such that $\lambda_q\leq\lambda_{q+1}$ for $q=1,2,\ldots,N_l$. The hidden layer is expanded in eigenmodes as $h=\sum_q z_qe_q$. In the microscopic model above, fluctuations of different hidden-layer eigenmodes satisfy $\langle z_q^2\rangle=1/(\beta\lambda_q)$. Hence, a smaller $\lambda_q$ corresponds to larger hidden-state fluctuations and thus a larger space of free variation, which can retain more information about the input and response.

We therefore consider a generation-$(l+1)$ student model whose hidden layer retains the low-eigenvalue modes of the teacher hidden layer, namely $\{\lambda_1,\lambda_2,\ldots,\lambda_{N_{l+1}}\}$. For this specific hidden-layer structure, according to Eq.~(\ref{eq:a(l)}), the relative HDC loss of the generation-$(l+1)$ model with respect to the generation-$l$ model is
\begin{equation}
a(l)=1-\frac{
\sum_{q=1}^{N_{l}}
u_{q}^{T}c_{q}/\lambda_{q}
}{
\sum_{q=1}^{N_{l-1}}
u_{q}^{T}c_{q}/\lambda_{q}
}, \label{eq:a(l)}
\end{equation}
Here, $u_q^T=u^Te_qe_q^T$ and $c_q=e_qe_q^Tc$ represent the projections of the hidden-layer-input coupling and the hidden-layer-output coupling onto the $q$th eigenmode $e_q$, respectively.

We next estimate the scaling behavior of $a(l)$. In the model above, for given input $x$ and response $y$, the hidden-layer variable $h$ obeys the conditional Gaussian distribution
\begin{equation}
    P(h|x, y)=\frac{1}{Z_{\mathrm{T}}(x, y)}\exp\left[-\frac{\beta}{2}(h-J^{-1}\eta)^T J(h-J^{-1}\eta)\right].\label{eq:Gaussian}
\end{equation}
Let the conditional mean be $\langle h\rangle_{x,y}=\int\mathrm{d}h\,P(h|x,y)h$. The conditional covariance matrix of the hidden layer is defined as
\begin{equation}
    \mathrm{Cov}(h|x, y)=\langle(h-\langle h\rangle_{x, y})(h-\langle h\rangle_{x, y})^T\rangle, 
\end{equation}
It follows from Eq.~(\ref{eq:Gaussian}) that $\mathrm{Cov}(h|x,y)=(\beta J)^{-1}$. Since $J$ has the eigenvalue relation $Je_q=\lambda_q e_q$, one has $\mathrm{Cov}(h|x,y)e_q=\sigma_q^2e_q$, where $\sigma_q^2$ is the eigenvalue of $\mathrm{Cov}(h|x,y)$ corresponding to the same eigenmode $e_q$ as that of $J$, with $\sigma_q^2=1/(\beta\lambda_q)$.

On the other hand, experiments \cite{agrawal2022alphareq} have found that, over a certain range, the covariance spectrum of neural-network hidden layers typically exhibits a power-law decay, $\sigma_q^2\sim q^{-\alpha}$, and hence $\lambda_q\sim q^\alpha$. Motivated by this observation, we assume that $u_q^Tc_q/\lambda_q$ approximately follows a power-law decay,
\begin{equation}
u_{q}^{T}c_{q}/\lambda_{q} \sim q^{-p}\leq N_l^{-p}, \qquad q\gg1
\end{equation}
Here, $p$ is the power-law exponent and the maximum value of $q$ is $N_l$. It is worth noting that conditional activation means and response-related quantities have been measured experimentally \cite{papyan2020neuralcollapse,arditi2024refusal}; these quantities correspond, respectively, to $J^{-1}u(h)$ and $J^{-1}c$. This indirectly indicates that the above power-law assumption can be tested experimentally.

Based on the above discussion, we now give the scaling behavior of $a(l)$ in different ranges of $p$. For $p>1$, $u_q^Tc_q/\lambda_q$ decays rapidly with $q$, so that
\begin{equation}
    \sum_{q=N_l}^{N_{l-1}}
u_{q}^{T}c_{q}/\lambda_{q}\ll\sum_{q=1}^{N_{l}}
u_{q}^{T}c_{q}/\lambda_{q}, 
\end{equation}
Therefore,
\begin{equation}
    a(l)=1-\frac{
\sum_{q=1}^{N_{l}}
u_{q}^{T}c_{q}/\lambda_{q}
}{
\sum_{q=1}^{N_{l}}
u_{q}^{T}c_{q}/\lambda_{q}+\sum_{q=N_l}^{N_{l-1}}
u_{q}^{T}c_{q}/\lambda_{q}
}\simeq 1-1-\frac{
\sum_{q=1}^{N_{l}}
u_{q}^{T}c_{q}/\lambda_{q}
}{
\sum_{q=1}^{N_{l}}
u_{q}^{T}c_{q}/\lambda_{q}
}=0.
\end{equation}
Thus, for $p>1$, the HDC of the student model is almost lossless relative to that of the teacher model.

For $p<1$, since $N_l,N_{l+1}\gg1$, one obtains
\begin{equation}
    a(l)\simeq 1-(1-s_l)^{1-p}, 
\end{equation}
It can be noted that when $s\ll1$, $a(l)\simeq(1-p)s$. According to the discussion in the main text, the critical alignment strengths satisfy $\gamma_\mathrm{PF},\gamma_\mathrm{tri}\propto a$. Therefore, these critical alignment strengths are proportional to the model-size reduction $s_l$.

Furthermore, when the same degree-of-freedom reduction ratio is adopted in every generation, namely $N_l=N_0(1-s)^l$, one has $a(l)\simeq1-(1-s)^{1-p}$, which is independent of $l$.

\section{Two-dimensional RG-like flow}
In this section, we derive the renormalization group like (RG-like) flow equations for the reliability-safety trade-off parameters under multigeneration distillation, namely Eq.~(18) of the main text, and analyze their fixed-point structure. Specifically, we first start from a single distillation step and obtain the changes of the trade-off parameters $A$ and $K$ over one generation of distillation. We then establish the two-dimensional RG-like flow equations in the continuous-generation limit. In Sec.~\ref{sec2.1_SM}, we first consider the case adopted in the main text, in which the HDC and overall response tendency of the student model before each generation of distillation satisfy $K_0=(1-a)K$ and $A_0=\mathrm{const}$, respectively. By solving the fixed-point equations, we systematically analyze the fixed-point structure and stability of the RG-like flow in different parameter regions under these conditions. In Sec.~\ref{sec2.2_SM}, we briefly demonstrate that the RG-like flow equations derived in the continuous-generation limit, as considered in the main text and Sec.~\ref{sec2.1_SM}, yield conclusions consistent with those obtained from the discrete-generation flow equations. In Sec.~\ref{sec2.3_SM}, we further discuss more general forms of the pre-distillation student-model parameters $K_0$ and $A_0$, showing that the fixed-point structures and bifurcation properties obtained in the main text are not restricted to the particular forms used in Sec.~\ref{sec2.1_SM}, but remain valid in a broader class of cases.

We first discuss a single distillation process. Let the HDC and overall refusal tendency of the student model before distillation be $K_0$ and $A_0$, respectively, and let the corresponding parameters of the teacher model be $K_\mathrm{T}$ and $A_\mathrm{T}$. According to Eqs.~(14) and (15) of the main text, the effective parameters of the student model after distillation satisfy
\begin{align}
K_{\mathrm{eff}}=&\,  K_{0}+B_{\mathrm{T}}(1)-B_{\mathrm{T}}(-1)\notag\\
=&\,  K_{0}+2\mathrm{arctanh}[\tanh\gamma(2R_{\mathrm{T}}-1)]+2\mathrm{arctanh}[\tanh\gamma(2S_{\mathrm{T}}-1)],  \\
A_{\mathrm{eff}}=&\,  A_{0}-\frac{1}{2\beta}[B_{\mathrm{T}}(1)+B_{\mathrm{T}}(-1)]\notag\\
=&\,  A_0-\frac{1}{\beta}\{\mathrm{arctanh}[\tanh\gamma(2R_{\mathrm{T}}-1)]-\mathrm{arctanh}[\tanh\gamma(2S_{\mathrm{T}}-1)]\}.
\end{align}
Here, $R_\mathrm{T}$ and $S_\mathrm{T}$ denote the reliability and safety of the teacher model, respectively. Under weak alignment, $\gamma\ll1$, the above two equations can be approximated as
\begin{align}
K_{\mathrm{eff}}\simeq&\,  K_{0}+4\gamma(R_{\mathrm{T}}+S_{\mathrm{T}}-1),  \label{eq:K_eff_SM}\\
A_{\mathrm{eff}}\simeq&\,  A_{0}+\frac{2\gamma}{\beta}(S_{\mathrm{T}}-R_{\mathrm{T}}).\label{eq:A_eff_SM}
\end{align}

We next consider multigeneration distillation, in which the generation-$l$ student model becomes the teacher model for the next generation after completing distillation. Let the reliability and safety of the generation-$l$ student model after distillation be $R(l)$ and $S(l)$, respectively, and let its HDC and overall response tendency be $K(l)$ and $A(l)$. The corresponding parameters of the generation-$(l+1)$ student model before distillation are denoted by $K_0(l+1)$ and $A_0(l+1)$. Under weak alignment, the intergenerational evolution relations follow from the single-step distillation results, Eqs.~(\ref{eq:K_eff_SM}) and (\ref{eq:A_eff_SM}). In addition, treating the generation index $l$ as a continuous variable, i.e., $K(l+1)-K(l)\rightarrow\mathrm{d}K(l)/\mathrm{d}l$ and $A(l+1)-A(l)\rightarrow\mathrm{d}A(l)/\mathrm{d}l$, we obtain the two-dimensional RG-like flow equations in the continuous-generation limit:
\begin{align}
\frac{\mathrm{d}K(l)}{\mathrm{d}l}=&\,  K_{0}(l+1)-K(l)+4\gamma\left[R(l)+S(l)-1\right]\notag\\=&\,  K_{0}(l+1)-K(l)+4\gamma\frac{\sinh\left[K(l)/2\right]}{\cosh\left[K(l)/2\right]+\cosh\left[\beta A(l)\right]},  \label{eqrg1}\\
\frac{\mathrm{d}A(l)}{\mathrm{d}l}=&\,  A_{0}(l+1)-A(l)+\frac{2\gamma}{\beta}\left[S(l)-R(l)\right]\notag\\=&\,  A_{0}(l+1)-A(l)+\frac{2\gamma}{\beta}\frac{\sinh\left[K(l)/2\right]}{\cosh\left[K(l)/2\right]+\cosh\left[\beta A(l)\right]}.\label{eqrg2}
\end{align}

\subsection{RG-like flow with a fixed HDC ratio and overall response tendency}\label{sec2.1_SM}
We first consider the case discussed in the main text: $K_0(l+1)=(1-a)K(l)$, i.e., the HDC loss of the pre-distillation student relative to its teacher is fixed in every generation. At the same time, the pre-distillation student model in every generation has the same overall refusal tendency, $A_0(l)\equiv A_0=\mathrm{const}$. Under these conditions, the two-dimensional RG-like flow in Eqs.~(\ref{eqrg1}) and (\ref{eqrg2}) becomes
\begin{align}
\frac{\mathrm{d}K}{\mathrm{d}l}
&\,  =-aK+4\gamma
\frac{\sinh(K/2)}
{\cosh(K/2)+\cosh(\beta A)}
\equiv\beta_{K}(A,  K),  \label{eq:finite-K-flow}\\
\frac{\mathrm{d}A}{\mathrm{d}l}
&=A_{0}-A+\frac{2\gamma}{\beta}
\frac{\sinh(\beta A)}
{\cosh(K/2)+\cosh(\beta A)}
\equiv\beta_{A}(A,  K).
\label{eq:finite-A-flow}
\end{align}

We now analyze the fixed-point structure and stability of this RG-like flow. A fixed point $(A_*,K_*)$ satisfies $\beta_K(A_*,K_*)=\beta_A(A_*,K_*)=0$, corresponding to a state in which the model parameters remain stable after many generations of distillation.

First consider the central fixed point $K_*=0$. Then $A_*$ satisfies
\begin{equation}
A_{*}=A_{0}+\frac{2\gamma}{\beta}
\tanh\left(\frac{\beta A_{*}}{2}\right).
\label{eq:finite-A-central}
\end{equation}
In the weak-alignment limit $\gamma\ll1$, retaining $\gamma$ to the lowest order gives
\begin{equation}
A_{*}=A_{0}+\frac{2\gamma}{\beta}
\tanh\left(\frac{\beta A_{0}}{2}\right)
+O(\gamma^{2}).
\end{equation}

We next discuss the stability of the central fixed point $(A_*,0)$. Its stability is determined by the eigenvalues of the Jacobian matrix at the fixed point,
\begin{equation}
\left.
\begin{pmatrix}
\dfrac{\partial\beta_{K}}{\partial K}&
\dfrac{\partial\beta_{K}}{\partial A}\\[6pt]
\dfrac{\partial\beta_{A}}{\partial K}&
\dfrac{\partial\beta_{A}}{\partial A}
\end{pmatrix}
\right|_{A=A_{*},  K=0}
=
\begin{pmatrix}
-a+\gamma\operatorname{sech}^{2}
\left(\dfrac{\beta A_{*}}{2}\right)&0\\[6pt]
0&-1+\gamma\operatorname{sech}^{2}
\left(\dfrac{\beta A_{*}}{2}\right)
\end{pmatrix}
\label{eq:finite-A-central-J}
\end{equation}
The two eigenvalues are
\begin{align}
    \lambda_K=&\,  -a+\gamma\operatorname{sech}^{2}
\left(\dfrac{\beta A_{*}}{2}\right)=-a+\gamma\operatorname{sech}^{2}
\left(\dfrac{\beta A_0}{2}\right)+O(\gamma^2),  \\
\lambda_A=&\,  -1+\gamma\operatorname{sech}^{2}
\left(\dfrac{\beta A_{*}}{2}\right)=-1+\gamma\operatorname{sech}^{2}
\left(\dfrac{\beta A_0}{2}\right)+O(\gamma^2).\label{eq:finite-A-pitchfork}
\end{align}
These correspond to the $K$ and $A$ directions, respectively. Since $\gamma\ll1$, one has $\lambda_A<0$, and hence the central fixed point is always stable in the $A$ direction. Its stability in the $K$ direction is determined by $\lambda_K$. When ${a}/{\gamma}
>\operatorname{sech}^{2}
\left({\beta A_0}/{2}\right)$, $\lambda_K<0$, and the central fixed point is a stable node, stable in both the $K$ and $A$ directions. Conversely, when ${a}/{\gamma}
<\operatorname{sech}^{2}
\left({\beta A_0}/{2}\right)$, the central fixed point is a saddle point and is unstable in the $K$ direction. When ${a}/{\gamma}
=\operatorname{sech}^{2}
\left({\beta A_0}/{2}\right)$, it is a nonhyperbolic fixed point, for which the linear stability criterion based on the Jacobian cannot determine the stability in the $K$ direction; nonlinear terms must be considered.

We next consider nonzero fixed points. Because the RG-like flow equations are invariant under the transformation $K\rightarrow-K$, nonzero fixed points always occur in pairs: if $(A_*',K_*')$, with $K_*'>0$, is a fixed point, then $(A_*',-K_*')$ is also a fixed point. Based on this property, we discuss only the positive fixed point $(A_*',K_*')$ below. The corresponding results for the negative fixed point $(A_*',-K_*')$ follow directly from $K_*'\rightarrow-K_*'$.

From Eq.~(\ref{eq:finite-A-flow}), the fixed-point equations $\beta_K(A_*',K_*')=\beta_A(A_*',K_*')=0$ can be written as
\begin{align}
\frac{a}{\gamma}
=&\,  
\frac{4\sinh(K'_{*}/2)}
{K'_{*}[\cosh(K'_{*}/2)+\cosh(\beta A_{*}')]},  \label{eq:finite-K-nonzero-exact}\\
A_{*}'=&\,  A_{0}+\frac{2\gamma}{\beta}
\frac{\sinh(\beta A_*')}
{\cosh(K_*'/2)+\cosh(\beta A_*')}.
\label{eq:finite-A-nonzero-exact}
\end{align}
In the weak-alignment limit $\gamma\ll1$, Eqs.~(\ref{eq:finite-A-nonzero-exact}) and (\ref{eq:finite-K-nonzero-exact}) can be approximated as
\begin{equation}
A_{*}'=A_{0}+\frac{2\gamma}{\beta}
\frac{\sinh(\beta A_{0})}
{\cosh(K'_{*}/2)+\cosh(\beta A_{0})}
+O(\gamma^{2}),  
\label{eq:finite-A-Ashift}
\end{equation}
and
\begin{equation}
\frac{a}{\gamma}
\simeq f(K_*')
,  
\label{eq:finite-A-root}
\end{equation}
where we define
\begin{equation}
f(K):=
\frac{4\sinh(K/2)}
{K[\cosh(K/2)+\cosh(\beta A_{0})]}.
\label{eq:fK}
\end{equation}
Thus, for a nonzero fixed point $(A_*',K_*')$, $K_*'$ is approximately determined by Eq.~(\ref{eq:finite-A-root}), while $A_*'$ is determined by Eq.~(\ref{eq:finite-A-Ashift}).

We next discuss the stability of the nonzero fixed point $(A_*',K_*')$. The Jacobian matrix at this fixed point is
\begin{equation}
\left.
\begin{pmatrix}
\dfrac{\partial\beta_{K}}{\partial K}&
\dfrac{\partial\beta_{K}}{\partial A}\\[6pt]
\dfrac{\partial\beta_{A}}{\partial K}&
\dfrac{\partial\beta_{A}}{\partial A}
\end{pmatrix}
\right|_{A=A_{*}',  K=K'_{*}}
=
\begin{pmatrix}
\displaystyle
-a+2\gamma
\frac{1+\cosh(\beta A_{*}')\cosh(K'_{*}/2)}
{[\cosh(K'_{*}/2)+\cosh(\beta A_{*}')]^{2}}
&
\displaystyle
-4\beta\gamma
\frac{\sinh(\beta A_{*}')\sinh(K'_{*}/2)}
{[\cosh(K'_{*}/2)+\cosh(\beta A_{*}')]^{2}}
\\[12pt]
\displaystyle
-\frac{\gamma}{\beta}
\frac{\sinh(\beta A_{*}')\sinh(K_{*}'/2)}
{[\cosh(K'_{*}/2)+\cosh(\beta A_{*}')]^{2}}
&
\displaystyle
-1+2\gamma
\frac{1+\cosh(\beta A_{*}')\cosh(K_{*}/2)}
{[\cosh(K_{*}/2)+\cosh(\beta A_{*}')]^{2}}
\end{pmatrix}.
\label{eq:finite-A-full-J}
\end{equation}
Its two eigenvalues in the weak-alignment limit are, respectively,
\begin{equation}
    \lambda_1=-1+2\gamma
\frac{1+\cosh(\beta A_0)\cosh(K'_{*}/2)}
{[\cosh(K'_{*}/2)+\cosh(\beta A_0)]^{2}}+O(\gamma^2),  \label{eq:lambda_1}
\end{equation}
and
\begin{equation}
    \lambda_2=\gamma K_*'f'(K_*')+O(\gamma^2).\label{eq:lambda_2}
\end{equation}
Equation (\ref{eq:lambda_1}) gives $\lambda_1=-1+O(\gamma)<0$, and therefore the nonzero fixed point $(A_*',K_*')$ is always stable along the direction corresponding to the eigenvalue $\lambda_1$. Stability along the direction corresponding to $\lambda_2$ is determined by the derivative $f'(K_*')$: if $f'(K_*')<0$, then $\lambda_2<0$ and the nonzero fixed point $(A_*',K_*')$ is stable; if $f'(K_*')>0$, then $\lambda_2>0$ and the nonzero fixed point is a saddle point, unstable along the direction corresponding to $\lambda_2$; if $f'(K_*')=0$, then $\lambda_2=0$ and the nonzero fixed point is nonhyperbolic.

According to the fixed-point equation (\ref{eq:finite-A-root}), the number of nonzero fixed points is determined by the number of intersections between $f(K)$ and $a/\gamma$. From the stability discussion above, the stability of the fixed points is mainly determined by the signs of $\lambda_K$ and $\lambda_2$. Since different values of the parameters $\beta A_0$ and $a/\gamma$ change the function $f(K)$ and the properties of $\lambda_K$ and $\lambda_2$, we next discuss the number and stability of nonzero fixed points in different regions of the parameter space $(\beta A_0,a/\gamma)$.

\subsubsection{$|\beta A_{0}|<\operatorname{arcosh}2$}\label{sec:betaA_0_lower}

We first discuss the region of relatively weak overall refusal tendency, $|\beta A_{0}|<\operatorname{arcosh}2$. In this region, $f(K)$ in Eq.~(\ref{eq:fK}) decreases monotonically for $K>0$ from $f(0)=\operatorname{sech}^2(\beta A_0/2)$ to $f(\infty)=0$. Therefore, $a/\gamma$ and $f(K)$ can have at most one intersection, i.e., Eq.~(\ref{eq:finite-A-root}) has at most one positive root, corresponding to a symmetric pair of nonzero fixed points $(A_*',\pm K_*')$. According to the number of intersections, the parameter range of $a/\gamma$ can be divided into the following three cases.

(i) $a/\gamma>f(0)$, i.e.,
\begin{equation}
\frac{a}{\gamma}>
\operatorname{sech}^{2}
\left(\frac{\beta A_{0}}{2}\right),  
\end{equation}
In this case, $a/\gamma$ does not intersect $f(K)$, and the system has no nonzero fixed point. Thus, the system has only the central fixed point $(A_*,0)$. In this parameter region, $\lambda_K<0$, so the central fixed point is a stable node.

(ii) $a/\gamma=f(0)$, i.e.,
\begin{equation}
\frac{a}{\gamma}=
\operatorname{sech}^{2}
\left(\frac{\beta A_{0}}{2}\right),  
\end{equation}
The system still has only the central fixed point. Since $\lambda_K=0$ in this region, the central fixed point is nonhyperbolic. The linear stability criterion based on the Jacobian matrix in Eq.~(\ref{eq:finite-A-central-J}) cannot determine the stability in the $K$ direction, so the nonlinear terms of the two-dimensional RG-like flow along the $K$ direction must be examined. Expanding the $K$-direction flow (\ref{eq:finite-K-flow}) near the central fixed point $K=0$ gives
\begin{equation}
    \frac{\mathrm{d}K}{\mathrm{d}l}=\frac{\cosh(\beta A_{0})-2}
{12[1+\cosh(\beta A_{0})]^{2}}K^3+O(K^5).\label{eq:dk_expansion}
\end{equation}
Since $|\beta A_0|<\operatorname{arcosh}(2)$, the coefficient of the cubic term is negative, and therefore the central fixed point remains nonlinearly stable.

(iii) $f(\infty)<a/\gamma<f(0)$, i.e.,
\begin{equation}
0<\frac{a}{\gamma}<
\operatorname{sech}^{2}
\left(\frac{\beta A_{0}}{2}\right),  
\end{equation}
In this case, $a/\gamma$ and $f(K)$ have one positive intersection, so Eq.~(\ref{eq:finite-A-root}) has one positive root. The system therefore has three fixed points: a symmetric pair of nonzero fixed points $(A_*',\pm K_*')$ and the central fixed point $(A_*,0)$. In this region, $\lambda_K>0$, so the central fixed point is a saddle and is unstable in the $K$ direction. On the other hand, $\lambda_2<0$, so the nonzero fixed points $(A_*',\pm K_*')$ are stable nodes.

The above discussion of the three $a/\gamma$ regions shows that, in the weak-tendency regime $|\beta A_0|<\operatorname{arcosh}2$, the system undergoes a supercritical pitchfork bifurcation: as $a/\gamma$ decreases from above $\operatorname{sech}^2
\left({\beta A_0}/{2}\right)$ to below it, the stable fixed point originally located at $K=0$ loses stability and a pair of stable nonzero fixed points $K=\pm K_*'$ is generated. Thus, ${a}/{\gamma}
=\operatorname{sech}^{2}
\left({\beta A_0}/{2}\right)$ corresponds to a supercritical pitchfork-bifurcation line, shown as the solid line to the left of the pentagram in Fig. 3 of the main text.

Near this supercritical bifurcation line, the scaling behavior of the nonzero fixed points $(A_*',\pm K_*')$ can be obtained from
\begin{equation}
   f(K_*')\simeq f(0)
+
\frac{\cosh(\beta A_{0})-2}
{12[1+\cosh(\beta A_{0})]^{2}}K_*'^{2}
,  
\label{eq:finite-A-small-K}
\end{equation}
and the fixed-point equation (\ref{eq:finite-A-root}), yielding
\begin{equation}
K_{*}\simeq
\sqrt{
\frac{12[1+\cosh(\beta A_{0})]^{2}}
{2-\cosh(\beta A_{0})}
\left[
\operatorname{sech}^{2}
\left(\frac{\beta A_{0}}{2}\right)
-\frac{a}{\gamma}
\right]
}.
\label{eq:finite-A-supercritical-root}
\end{equation}
Therefore, at fixed $a$ and $\beta A_0$, when $\gamma$ is varied, the nonzero fixed point near the bifurcation line satisfies the square-root scaling relation
\begin{equation}
    K_*'\propto|\gamma-\gamma_{\mathrm{PF}}|^{1/2},  
\end{equation}
where $\gamma_\mathrm{PF}=a\cosh^2(\beta A_0/2)$.

\subsubsection{$|\beta A_0|>\operatorname{arcosh}2$}

We now discuss the region of relatively strong overall response tendency, $|\beta A_0|>\operatorname{arcosh}2$. In this region, the function $f(K)$ in Eq.~(\ref{eq:fK}) first increases and then decreases for $K>0$. Let the maximum of $f(K)$ occur at $K_m>0$, satisfying $f'(K_m)=0$. This gives
\begin{equation}
\frac{K_{\mathrm{m}}}{2}
\left[
1+\cosh(\beta A_{0})
\cosh\left(\frac{K_{\mathrm{m}}}{2}\right)
\right]
=
\sinh\left(\frac{K_{\mathrm{m}}}{2}\right)
\left[
\cosh\left(\frac{K_{\mathrm{m}}}{2}\right)
+\cosh(\beta A_{0})
\right].
\label{eq:finite-A-maximum}
\end{equation}
Thus, in the region $|\beta A_0|>\operatorname{arcosh}2$, as $K$ increases, $f(K)$ first increases monotonically from $f(0)=\operatorname{sech}^2(\beta A_0/2)$ to $f(K_m)$ and then decreases monotonically to $f(\infty)=0$. Hence, $a/\gamma$ and $f(K)$ can have at most two intersections, so Eq.~(\ref{eq:finite-A-root}) has at most two positive roots, corresponding to two pairs of nonzero fixed points. According to the intersection positions, the parameter $a/\gamma$ can be divided into the following five cases.

(i) $a/\gamma>f(K_m)$, i.e.,
\begin{equation}
\frac{a}{\gamma}>
\frac{4\sinh(K_{\mathrm{m}}/2)}
{K_{\mathrm{m}}
[\cosh(K_{\mathrm{m}}/2)+\cosh(\beta A_{0})]},  
\end{equation}
Equation (\ref{eq:finite-A-root}) has no positive root, so the system contains only the central fixed point $(A_*,0)$. In this region, $\lambda_A,\lambda_K<0$, and the fixed point is a stable node.

(ii) $a/\gamma=f(K_m)$, i.e.,
\begin{equation}
\frac{a}{\gamma}=
\frac{4\sinh(K_{\mathrm{m}}/2)}
{K_{\mathrm{m}}
[\cosh(K_{\mathrm{m}}/2)+\cosh(\beta A_{0})]}.
\end{equation}
Equation (\ref{eq:finite-A-root}) has one positive root, and hence the system has three fixed points: a symmetric pair of nonzero fixed points $(A_*',\pm K_m)$ and the central fixed point $(A_*,0)$. In this region, $\lambda_K<0$, so the central fixed point is a stable node; meanwhile, $\lambda_2=0$, so the nonzero fixed points $(A_*',\pm K_m)$ are nonhyperbolic.

(iii) $f(0)<a/\gamma<f(K_m)$, i.e.,
\begin{equation}
\operatorname{sech}^{2}
\left(\frac{\beta A_{0}}{2}\right)
<
\frac{a}{\gamma}
<
\frac{4\sinh(K_{\mathrm{m}}/2)}
{K_{\mathrm{m}}
[\cosh(K_{\mathrm{m}}/2)+\cosh(\beta A_{0})]},  
\end{equation}
Equation (\ref{eq:finite-A-root}) has two positive roots, denoted by $K_{*,\mathrm{in}}$ and $K_{*,\mathrm{out}}$, with $0<K_{*,\mathrm{in}}<K_m<K_{*,\mathrm{out}}$. Thus, the system has five fixed points: the central fixed point $(A_*,0)$, and two pairs of nonzero fixed points $(A_{*,\mathrm{in}}',\pm K_{*,\mathrm{in}})$ and $(A_{*,\mathrm{out}}',\pm K_{*,\mathrm{out}})$. In this region, $\lambda_K<0$, so the central fixed point is a stable node. For the inner fixed points $(A_{*,\mathrm{in}}',\pm K_{*,\mathrm{in}})$, $f'(K_{*,\mathrm{in}})>0$, and according to Eq.~(\ref{eq:lambda_2}), $\lambda_{2,\mathrm{in}}=\gamma K_{*,\mathrm{in}}f'(K_{*,\mathrm{in}})>0$, so they are saddle points. For the outer fixed points $(A_{*,\mathrm{out}}',\pm K_{*,\mathrm{out}})$, $\lambda_{2,\mathrm{out}}=\gamma K_{*,\mathrm{out}}f'(K_{*,\mathrm{out}})<0$, so they are stable nodes.

(iv) $a/\gamma=f(0)$, i.e.,
\begin{equation}
\frac{a}{\gamma}=
\operatorname{sech}^{2}
\left(\frac{\beta A_{0}}{2}\right),  
\end{equation}
The system then has only three fixed points: the central fixed point $(A_*,0)$ and one outer pair of nonzero fixed points $(A_{*,\mathrm{out}}',\pm K_{*,\mathrm{out}})$. In this region, $\lambda_{2,\mathrm{out}}<0$ for the outer fixed-point pair, so these are stable nodes. For the central fixed point, $\lambda_K=0$, so it is nonhyperbolic and its nonlinear terms in the $K$ direction must be examined. According to Eq.~(\ref{eq:dk_expansion}), under the condition $|\beta A_0|>\operatorname{arcosh}(2)$ the coefficient of the cubic term in Eq.~(\ref{eq:dk_expansion}) is positive; the central fixed point is therefore nonlinearly unstable in the $K$ direction.

(v) $f(\infty)<a/\gamma<f(0)$, i.e.,
\begin{equation}
0<\frac{a}{\gamma}<
\operatorname{sech}^{2}
\left(\frac{\beta A_{0}}{2}\right),  
\end{equation}
The system has three fixed points: the central fixed point $(A_*,0)$ and one outer pair of nonzero fixed points $(A_{*,\mathrm{out}}',\pm K_{*,\mathrm{out}})$. In this region, $\lambda_{2,\mathrm{out}}<0$ for the outer fixed-point pair, so they are stable nodes. Since $\lambda_2>0$, the central fixed point is a saddle point.

The above discussion of the five $a/\gamma$ regions shows that, in the strong-tendency regime $|\beta A_0|>\operatorname{arcosh}2$, the system exhibits two different types of bifurcations. First, when $a/\gamma$ decreases from above $f(K_m)$ to below it, the system generates two pairs of nonzero fixed points. At $a/\gamma=f(K_m)$, the two pairs $\pm K_{*,\mathrm{in}}$ and $\pm K_{*,\mathrm{out}}$ merge at $\pm K_m$ and are nonhyperbolic. Therefore, $a/\gamma=f(K_m)$ corresponds to a saddle-node bifurcation line, shown as the dashed line in Fig. 3 of the main text. As $a/\gamma$ decreases further to $f(0)$, the inner nonzero fixed points $\pm K_{*,\mathrm{in}}$ connect with the central fixed point $K=0$. Because these fixed points are all saddle points, in the region $|\beta A_0|>\operatorname{arcosh}2$, ${a}/{\gamma}
=\operatorname{sech}^{2}
\left({\beta A_0}/{2}\right)$ corresponds to a subcritical pitchfork-bifurcation line, shown as the solid line to the right of the pentagram in Fig.~4 of the main text.

Near the subcritical pitchfork-bifurcation line ${a}/{\gamma}
=\operatorname{sech}^{2}
\left({\beta A_0}/{2}\right)$, the inner nonzero fixed points $(A_*',\pm K_{*,\mathrm{in}})$ satisfy
\begin{equation}
K_{*,  \mathrm{in}}\simeq
\sqrt{
\frac{12[1+\cosh(\beta A_{0})]^{2}}
{\cosh(\beta A_{0})-2}
\left[
\frac{a}{\gamma}
-\operatorname{sech}^{2}
\left(\frac{\beta A_{0}}{2}\right)
\right]
}.
\label{eq:finite-A-subcritical-root}
\end{equation}
Therefore, at fixed $a$ and $\beta A_0$, when $\gamma$ is varied, the inner nonzero fixed points $\pm K_{*,\mathrm{in}}$ near the bifurcation line satisfy the square-root scaling relation
\begin{equation}
   K_{*,  \mathrm{in}}\propto|\gamma-\gamma_{\mathrm{PF}}|^{1/2},  
\end{equation}
where $\gamma_\mathrm{PF}=a\cosh^2(\beta A_0/2)$.

\subsubsection{$|\beta A_{0}|=\operatorname{arcosh}2$}
We now discuss the critical region $|\beta A_{0}|=\operatorname{arcosh}2$. In this case, $f(K)$ in Eq.~(\ref{eq:fK}) decreases monotonically for $K>0$ from $f(0)=2/3$ to $f(\infty)=0$, giving the same fixed-point structure as in Sec. \ref{sec:betaA_0_lower}. Hence, the parameter $a/\gamma$ can be divided into the following three regions.

(i) $a/\gamma>f(0)=2/3$. The system contains only the central fixed point $(A_*,0)$, which is a stable node.

(ii) $a/\gamma=2/3$. The system still contains only the central fixed point $(A_*,0)$. Since this is a nonhyperbolic fixed point, the nonlinear terms in the $K$ direction must be further examined. Expanding the $K$-direction flow (\ref{eq:finite-K-flow}) near $K=0$ gives
\begin{equation}
\left.
\frac{\mathrm{d}K}{\mathrm{d}l}
\right|_{a/\gamma=2/3}
=
-\frac{\gamma}{4320}K^{5}
+O(K^{7}),  
\end{equation}
Since the coefficient of the fifth-order term is negative, the central fixed point remains nonlinearly stable.

(iii) $0<a/\gamma<f(0)=2/3$. The system has three fixed points: a symmetric pair of nonzero fixed points $(A_*',\pm K_*')$ and the central fixed point $(A_*,0)$. The central fixed point is a saddle, whereas the nonzero fixed points $(A_*',\pm K_*')$ are stable nodes.

At the point $(|\beta A_0|,a/\gamma)=(\operatorname{arcosh}2,2/3)$, the above analysis shows that the subcritical pitchfork-bifurcation line ${a}/{\gamma}
=\operatorname{sech}^{2}
\left({\beta A_0}/{2}\right)$ and the saddle-node bifurcation line $a/\gamma=f(K_m)$ from the region $|\beta A_0|>\operatorname{arcosh}2$, together with the supercritical pitchfork-bifurcation line ${a}/{\gamma}
=\operatorname{sech}^{2}
\left({\beta A_0}/{2}\right)$ from the region $|\beta A_{0}|<\operatorname{arcosh}2$, meet at this point. Moreover, this point simultaneously joins the regions with one, three, and five fixed points, and is therefore a tricritical point, as indicated by the pentagram in Fig.~4 of the main text.

Near the tricritical point, the nonzero fixed points $\pm K_*'$ satisfy
\begin{equation}
    K_*'\simeq \left[4320\left(\frac{2}{3}-\frac{a}{\gamma}\right)\right]^{1/4},  
\end{equation}
Therefore, at fixed $a$ and $\beta A_0$, when $\gamma$ is varied, the nonzero fixed point $K_*'$ near the three-phase point satisfies the fourth-root scaling relation
\begin{equation}
    K_*'\propto|\gamma-\gamma_{\mathrm{tri}}|^{1/4},  
\end{equation}
where $\gamma_\mathrm{tri}=3a/2$.

\subsection{Consistency with the discrete generational flow equations}\label{sec2.2_SM}
The continuous flow originates from the discrete generation-by-generation iteration. For the setting of the main text, $K^0_{l+1}=(1-a)K_l$ and $A^0_{l+1}=A_0$, the discrete iteration equations in the small-$\gamma$ limit are
\begingroup
\renewcommand{\theHequation}{SM.manual.S75}
\begin{equation}
\begin{pmatrix}
K_{l+1}-K_l\\
A_{l+1}-A_l
\end{pmatrix}
=
\begin{pmatrix}
\dfrac{4\gamma\sinh\frac{K_l}{2}}
{\cosh\frac{K_l}{2}+\cosh\beta A_l}
-aK_l
\\[8pt]
\dfrac{2\gamma\sinh\beta A_l}
{\beta\left(\cosh\frac{K_l}{2}+\cosh\beta A_l\right)}
+A_0-A_l
\end{pmatrix}
+O(\gamma^3).
\tag{S75}
\end{equation}
\endgroup

To the order retained, the weak-alignment discrete map and the continuous flow have the same fixed-point equations. Therefore, the fixed points of the discrete equations coincide with those in the continuous case.

We next need to show that their stability is also fully consistent, which requires examining the Jacobian of the above equations at a fixed point. The off-diagonal elements of the Jacobian $J_\beta$ are small quantities of order $O(\gamma)$:
\begin{equation}
J_\beta(K, A)
=
\begin{pmatrix}
\dfrac{\mathrm d}{\mathrm dK}
\dfrac{4\gamma\sinh\frac{K}{2}}
{\cosh\frac{K}{2}+\cosh\beta A}
-a
&
O(\gamma)
\\[10pt]
O(\gamma)
&
\dfrac{\mathrm d}{\mathrm dA}
\dfrac{2\gamma\sinh\beta A}
{\beta\left(\cosh\frac{K}{2}+\cosh\beta A\right)}
-1
\end{pmatrix}.
\end{equation}

For the above equations, denote the Jacobian eigenvalues by $\chi_i$, $i=1,2$. To $O(\gamma)$, the eigenvalues correspond to the two diagonal elements of the above matrix:
\begin{equation}
\chi_1
=
\frac{\mathrm d}{\mathrm dK}
\frac{4\gamma\sinh\frac{K}{2}}
{\cosh\frac{K}{2}+\cosh\beta A}
-a
+O(\gamma^2), 
\end{equation}

\begin{equation}
\chi_2
=
\frac{\mathrm d}{\mathrm dA}
\frac{2\gamma\sinh\beta A}
{\beta\left(\cosh\frac{K}{2}+\cosh\beta A\right)}
-1
+O(\gamma^2).
\end{equation}

The conditions $|1+\chi_1|<1$ and $|1+\chi_2|<1$ are the stability conditions of the fixed point of the discrete iteration against perturbations in the $K$ and $A$ directions, respectively. For the continuous iteration, $\chi_1<0$ and $\chi_2<0$ are the corresponding stability conditions. Therefore, if a fixed point is unstable in a given direction under the continuous iteration, it is necessarily unstable in the discrete case as well. The consistency check thus only needs to focus on directions in which a fixed point is stable.

We first examine stability in the $A$ direction. Since
\begin{equation}
\left|
\frac{\mathrm d}{\mathrm dA}
\frac{2\gamma\sinh\beta A}
{\beta\left(\cosh\frac{K}{2}+\cosh\beta A\right)}
\right|
\leq \gamma, 
\end{equation}
one has $-2<\chi_2<0$ for small $\gamma$. This implies $|1+\chi_2|<1$, i.e., in the discrete case, any fixed point is stable against perturbations in the $A$ direction, consistent with the continuous case.

We next examine stability in the $K$ direction. From the previous analysis, fixed points occur only when $a=O(\gamma)$. Therefore, for every stable fixed point of the continuous iteration, one has $\chi_1<0$ and $\chi_1=O(\gamma)$. Hence the condition $|\chi_1+1|<1$ is satisfied, so the fixed point remains stable under the discrete iteration.

In summary, the locations of the fixed points and their stability in each direction are identical for the discrete and continuous iterations.

\subsection{Discussion of the more general case}\label{sec2.3_SM}
In the main text, we considered the simple case $K_0(l+1)=(1-a)K(l)$ and $A_0(l)=\mathrm{const}$, i.e., the pre-distillation student model has a fixed HDC loss relative to the teacher model in every generation and the same overall response tendency. We now show that the fixed-point structures, bifurcation properties, and corresponding phase diagram obtained in the main text are not restricted to this particular form, but can remain valid in a broader class of cases. We briefly discuss the conditions required for these conclusions to hold.

Consider more general pre-distillation student-model HDC and overall response tendency, related to the corresponding teacher-model parameters by
\begin{equation}
    K_0(l+1)=g(K(l)),  \qquad A_0(l+1)=h(A(l)).
\end{equation}
The two-dimensional RG-like flows in Eqs.~(\ref{eqrg1}) and (\ref{eqrg2}) then become
\begin{align}
\frac{\mathrm{d}K}{\mathrm{d}l}
&\,  =g(K)-K+4\gamma
\frac{\sinh(K/2)}
{\cosh(K/2)+\cosh(\beta A)}
\equiv\tilde\beta_{K}(A,  K),  \label{eq:robust-general-flowK}\\
\frac{\mathrm{d}A}{\mathrm{d}l}
&=h(A)-A+\frac{2\gamma}{\beta}
\frac{\sinh(\beta A)}
{\cosh(K/2)+\cosh(\beta A)}
\equiv\tilde\beta_{A}(A,  K).
\label{eq:robust-general-flowA}
\end{align}
Thus, the case considered in the main text corresponds to the special choices $g(K)=(1-a)K$ and $h(A)=A_0=\mathrm{const}$. We now give, in turn, the requirements on $h(A)$ and $g(K)$ under which the phase diagram obtained in the main text remains valid.

First, the requirements on $h(A)$ are:

(i) $h(A)$ has a stable fixed point $\bar A$, satisfying $\bar A=h(\bar A)$.

(ii) $h(A)$ is such that $\tilde\beta_A(A,K)$ has a stable fixed point $A_*(K,\gamma)$ satisfying $A_*-\bar A=O(\gamma)$, i.e., the two fixed points differ only by a quantity of order $\gamma$.

It then follows that the fixed point of $\tilde\beta_A(A,K)$ is
\begin{equation}
A_{\mathrm{*}}(K;\gamma)
=\bar A+
\frac{2\gamma}{\beta[1-h'(\bar A)]}
\frac{\sinh(\beta\bar A)}
{\cosh(K/2)+\cosh(\beta\bar A)}
+O(\gamma^{2}).
\label{eq:robust-As}
\end{equation}
It can be noted that its dependence on $K$ is weak. We now focus on the long-time evolution $l\rightarrow\infty$ of $K$. Setting $A\simeq A_*$ and substituting into Eq.~(\ref{eq:robust-general-flowK}), one obtains the reduced one-dimensional flow in the $K$ direction,
\begin{equation}
\frac{\mathrm{d}K}{\mathrm{d}l}
=g(K)-K+4\gamma
\frac{\sinh(K/2)}
{\cosh(K/2)+\cosh(\beta\bar A)}
+O(\gamma^{2}).
\label{eq:robust-reduced-flow}
\end{equation}
Expanding the right-hand side about $K=0$ to fifth order, with $g(K)=g_1K+g_2K^2+g_3K^3+g_4K^4+g_5K^5+O(K^7)$, gives
\begin{equation}
\frac{\mathrm{d}K}{\mathrm{d}l}
=rK+g_2 K^2+uK^{3}+g_4 K^4+vK^{5}+O(K^{7},  \gamma^{2}),  
\label{eq:robust-landau}
\end{equation}
where, defining $C=\cosh(\beta\bar A)$,
\begin{align}
r&=g_{1}-1+\frac{2\gamma}{1+C}+O(\gamma^{2}),  \nonumber\\
u&=g_{3}+\frac{\gamma(C-2)}{12(1+C)^{2}}
+O(\gamma^{2}),  \nonumber\\
v&=g_{5}+\frac{\gamma(C^{2}-13C+16)}
{960(1+C)^{3}}+O(\gamma^{2}).
\label{eq:robust-landau-coefficients}
\end{align}
When $h(A)=A_0$ (i.e., $\bar A=A_0$) and $g(K)=(1-a)K$, Eq.~(\ref{eq:robust-landau}) reduces to the coefficients used in the main text.

We further impose the following requirements on $g(K)$:

(i) $g(K)$ is an odd function, i.e., $g_2=g_4=0$.

(ii) There exist values of the parameters $C$ and $\gamma$ such that
\begin{equation}
    r=0,  \qquad u=0,  \qquad v<0,  
\end{equation}
Under these conditions, the RG-like flow retains the fifth-order Ginzburg-Landau structure of the phase diagram in the main text. Consequently, the fixed-point structure, bifurcation properties, and phase-diagram structure obtained in the main text remain unchanged.

\section{Experimental interpretation of the HDC relation}

In this section, we establish a correspondence between the experimentally observed quantities reported in the refusal-token experiment of Ref.~\cite{jain2024refusal} and the macroscopic variables in the binary model of this work, and use the experimental data to estimate the HDC (hazard discrimination capability) in the model of the main text. The purpose of this section is to test a central prediction of the main text: when only the overall response tendency $A$ is changed, a fixed model moves along a reliability-safety trade-off curve with approximately constant HDC $K$.

\subsection{Correspondence between experimental observables and theoretical variables}
The experiment is based on a trained Llama 3 8B model. For each input query, the model first probabilistically generates a refusal token: before determining whether the input is safe, the model directly refuses. The probability of generating the refusal token can be adjusted by the model threshold parameter $T$. The model then uses the semantic information of the input query to determine whether it is safe and responds by either answering or refusing. Thus, the threshold $T$ controls the overall response tendency of the model and corresponds to the parameter $A$ in the main text. The true-positive rate (TPR) and false-positive rate (FPR) of semantically determined responses correspond to the reliability and safety in the main text.
\begin{table*}[t]
\centering
\begin{tabular}{|l|l|l|}
\hline
\tcell{Experimental object} & \tcell{Experimental observable} & \tcell{Binary-model quantity} \\ \hline
\tcell{Query that should be refused} & \tcell{Positive evaluation class} & \tcell{Hazardous input, $x=-1$} \\ \hline
\tcell{Answerable contrast query} & \tcell{Negative evaluation class} & \tcell{Ordinary input, $x=+1$} \\ \hline
\tcell{Semantically judged refusal} & \tcell{Refusal outcome} & \tcell{$y=-1$} \\ \hline
\tcell{Semantically judged answer} & \tcell{Response outcome} & \tcell{$y=+1$} \\ \hline
\tcell{Correct refusal rate} & \tcell{$\mathrm{TPR}=P(\text{refusal}\mid\text{refusal query})$} & \tcell{$S=P(y=-1\mid x=-1)$} \\ \hline
\tcell{Incorrect refusal rate} & \tcell{$\mathrm{FPR}=P(\text{refusal}\mid\text{contrast query})$} & \tcell{$1-R=P(y=-1\mid x=+1)$} \\ \hline
\tcell{Refusal-token score and threshold} & \tcell{$p_{\mathrm{refuse}}$ and $T$} & \tcell{Internal readout and a control of the overall response tendency $A$; neither is directly equal to $S$ or $K$} \\ \hline
\tcell{Training configuration} & \tcell{With or without contrast data} & \tcell{May change the input-response separation $q$, and hence $K=2\beta q$} \\ \hline
\end{tabular}
\caption{\label{tab:experimental_mapping}
Mapping between the refusal-token experiment and the binary HDC model.}
\end{table*}

We first give the correspondence between the measured objects in the experiment and the theoretical variables of this work, as summarized in Table \ref{tab:experimental_mapping}. The experiment divides the inputs into ``answerable contrast queries'' and ``queries that should be refused'', corresponding respectively to the safe and hazardous inputs in this work. In addition, the experimentally measured true-positive rate TPR and false-positive rate FPR denote, respectively, the probability that the model refuses a ``query that should be refused'' and the probability that it refuses an ``answerable contrast query.'' According to our definitions, the reliability and safety of the experimental model can be written as
\begin{equation}
R=1-\mathrm{FPR}, 
\qquad
S=\mathrm{TPR}.
\label{eq:experimental_RS}
\end{equation}

We next derive the relation between TPR and FPR from the results of this work. From the reliability-safety trade-off relation
\begin{equation}
    \mathrm{logit}R+\mathrm{logit} S=K, \label{eq:hdc_trade_off}
\end{equation}
one obtains
\begin{equation}
    \mathrm{logit}(\mathrm{TPR})-\mathrm{logit}(\mathrm{FPR})=K, 
\end{equation}
or equivalently,
\begin{equation}
    \mathrm{TPR}\left(\mathrm{FPR}, K\right)=\frac{e^K\mathrm{FPR}}{1-\mathrm{FPR}+e^K\mathrm{FPR}}, 
\end{equation}
where $K$ measures the capability of different models to discriminate semantically whether an input is safe.

\subsection{Agreement between the experimental data and theoretical predictions}
We now use the data from Ref.~\cite{jain2024refusal} to test the conclusions of this work. Since the threshold $T$ of the refusal token adjusts only the overall response tendency of the model, i.e., it changes only the parameter $A$, while our results show that $K$ is independent of $A$, we predict that varying $T$ does not change the HDC parameter $K$.

\begin{figure}
    \centering
    \includegraphics[width=0.8\linewidth]{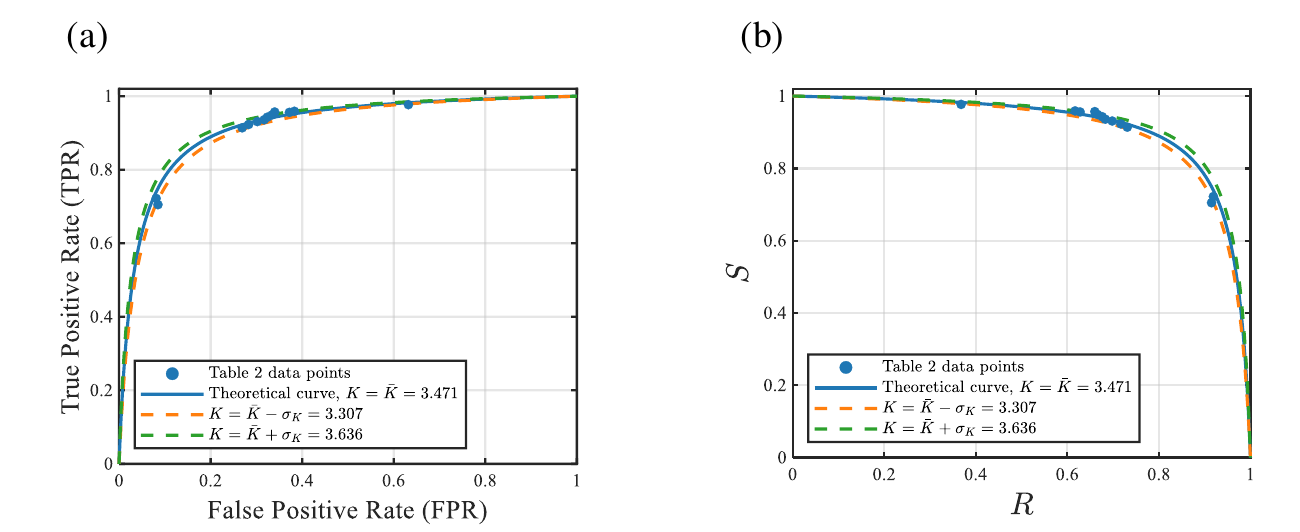}
    \caption{$S$--$R$ and TPR--FPR plots for the data in Table~2. The solid line represents the reliability--safety trade-off curve corresponding to the averaged HDC $\bar{K}=3.471$ obtained from all data points. The two dashed lines represent the trade-off curves with $K=\bar{K}\pm\sigma_K$, where $\sigma_K=0.164$ denotes the standard deviation of the HDC extracted from different experimental points.}
    \label{fig:table2_figure}
\end{figure}

\begin{figure}
    \centering
    \includegraphics[width=0.8\linewidth]{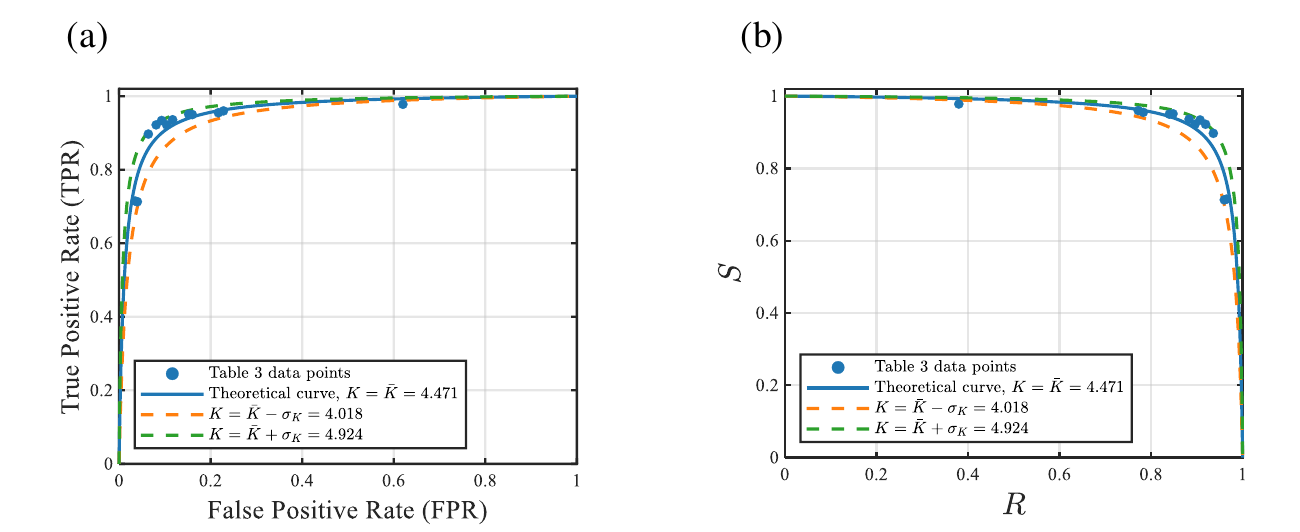}
    \caption{$S$--$R$ and TPR--FPR plots for the data in Table~2. The solid line represents the reliability--safety trade-off curve corresponding to the averaged HDC $\bar{K}=4.471$ obtained from all data points. The two dashed lines represent the trade-off curves with $K=\bar{K}\pm\sigma_K$, where $\sigma_K=0.453$ denotes the standard deviation of the HDC extracted from different experimental points.}
    \label{fig:table3_figure}
\end{figure}

To verify this theoretical prediction, we digitized representative data points from Fig.~2 of Ref.~\cite{jain2024refusal}. Specifically, Fig.~2 presents the TPR and FPR values corresponding to different refusal thresholds $T$ as discrete data points connected by line segments, forming piecewise linear curves. We extracted the corresponding TPR and FPR values from the endpoints of these line segments and calculated the corresponding $K$ values at different thresholds $T$ according to Eq.~(\ref{eq:hdc_trade_off}). The results are shown in Tables \ref{table2} and \ref{tabel3}, corresponding, respectively, to the CoCoNot model trained without contrast data and that trained with contrast data. The index $i=1,\ldots,12$ labels the data at different thresholds $T$. Although the TPR and FPR vary substantially across thresholds, the calculated $K$ remains approximately stable. Similarly, Figs.~\ref{fig:table2_figure} and \ref{fig:table3_figure} show the $(R,S)$ and $(\mathrm{FPR},\mathrm{TPR})$ data from Tables~\ref{table2} and \ref{tabel3}, respectively. This is approximately consistent with the theoretical prediction that changing the overall response tendency $A$ leaves the HDC parameter $K$ unchanged.

\begin{table*}[t]
\centering
\begin{tabular}{|c|c|c|c|c|}
\hline
$i$
& $S_i=\mathrm{TPR}_i$
& $1-R_i=\mathrm{FPR}_i$
& $R_i$
& $K_i$
\\ \hline
1  & 0.977 & 0.632 & 0.368 & 3.208 \\ \hline
2  & 0.959 & 0.383 & 0.617 & 3.629 \\ \hline
3  & 0.956 & 0.372 & 0.628 & 3.602 \\ \hline
4  & 0.957 & 0.340 & 0.660 & 3.766 \\ \hline
5  & 0.949 & 0.334 & 0.666 & 3.614 \\ \hline
6  & 0.943 & 0.324 & 0.676 & 3.541 \\ \hline
7  & 0.936 & 0.317 & 0.683 & 3.450 \\ \hline
8  & 0.931 & 0.302 & 0.698 & 3.440 \\ \hline
9  & 0.923 & 0.283 & 0.717 & 3.413 \\ \hline
10 & 0.914 & 0.269 & 0.731 & 3.363 \\ \hline
11 & 0.722 & 0.081 & 0.919 & 3.383 \\ \hline
12 & 0.705 & 0.085 & 0.915 & 3.247 \\ \hline
\end{tabular}
\caption{\label{table2} HDC evaluated at the twelve digitized vector vertices of the
threshold-sweep curve for the CoCoNot model trained without contrast
data. No averaging over the digitized vertices is performed.}
\end{table*}

\begin{table*}[t]
\centering
\begin{tabular}{|c|c|c|c|c|}
\hline
$i$
& $S_i=\mathrm{TPR}_i$
& $1-R_i=\mathrm{FPR}_i$
& $R_i$
& $K_i$
\\ \hline
1 & 0.978 & 0.620 & 0.380 & 3.305 \\ \hline
2 & 0.960 & 0.228 & 0.772 & 4.398 \\ \hline
3 & 0.955 & 0.217 & 0.783 & 4.338 \\ \hline
4 & 0.950 & 0.160 & 0.840 & 4.603 \\ \hline
5 & 0.951 & 0.152 & 0.848 & 4.685 \\ \hline
6 & 0.936 & 0.117 & 0.883 & 4.704 \\ \hline
7 & 0.934 & 0.093 & 0.907 & 4.927 \\ \hline
8 & 0.922 & 0.104 & 0.896 & 4.623 \\ \hline
9 & 0.922 & 0.081 & 0.919 & 4.899 \\ \hline
10 & 0.897 & 0.064 & 0.936 & 4.847 \\ \hline
11 & 0.715 & 0.035 & 0.965 & 4.237 \\ \hline
12 & 0.713 & 0.040 & 0.960 & 4.088 \\ \hline
\end{tabular}
\caption{\label{tabel3} HDC evaluated at the twelve digitized vector vertices of the
threshold-sweep curve for the CoCoNot model trained with contrast data.
No averaging over the digitized vertices is performed.}
\end{table*}

We further compare the models trained with and without contrast data. The purpose of contrast training is to improve the model's ability to distinguish hazardous inputs, so it should theoretically correspond to a larger HDC $K$. As shown in Tables \ref{table2} and \ref{tabel3}, the $K$ values extracted for the model trained with contrast data are systematically higher than those for the model trained without contrast data, further supporting the theoretical interpretation of HDC in this work.

\clearpage

\end{document}